\documentclass[sigconf]{acmart}
\usepackage{listings}
\usepackage{amssymb}
\usepackage{amsfonts}
\usepackage{amsmath}
\usepackage{graphicx}
\usepackage{nomencl}
\usepackage{subcaption}
\usepackage{breakurl}
\usepackage{booktabs}
\usepackage{subfiles}
\usepackage{amsthm}
\theoremstyle{definition}
\usepackage{tikz, pgfplots}
\usepackage{mathtools}
\renewcommand\footnotetextcopyrightpermission[1]{} 
\usetikzlibrary{matrix,positioning,arrows,shapes.gates.logic.US,shapes.gates.logic.IEC,calc}
\pgfplotsset{compat=newest}
\definecolor{tudelft}{RGB}{0,166,214}
\renewcommand\footnotetextcopyrightpermission[1]{} 
\setcopyright{none}
\definecolor{darkgreen}{rgb}{0,0.5,0}

\tikzset{ 
	table/.style={
		matrix of nodes,
		row sep=-\pgflinewidth,
		column sep=-\pgflinewidth,
		nodes={rectangle,draw=black,text width=1.5ex,align=center},
		text depth=0.25ex,
		text height=1ex,
		nodes in empty cells
	},
	texto/.style={font=\footnotesize\sffamily},
	title/.style={font=\small\sffamily}
}

\tikzset{ 
	tablemiddle/.style={
		matrix of nodes,
		row sep=-\pgflinewidth,
		column sep=-\pgflinewidth,
		nodes={rectangle,draw=black,text width=2ex,align=center},
		text depth=0.25ex,
		text height=1ex,
		nodes in empty cells
	},
	texto/.style={font=\footnotesize\sffamily},
	title/.style={font=\small\sffamily}
}

\tikzset{ 
	tablesmall/.style={
		matrix of nodes,
		row sep=-\pgflinewidth,
		column sep=-\pgflinewidth,
		nodes={rectangle,draw=black,text width=0.5ex,align=center},
		text depth=0.01ex,
		text height=0.5ex,
		nodes in empty cells
	},
	texto/.style={font=\footnotesize\sffamily},
	title/.style={font=\small\sffamily}
}

\tikzset{ 
	tablelong/.style={
		matrix of nodes,
		row sep=-\pgflinewidth,
		column sep=-\pgflinewidth,
		nodes={rectangle,draw=black,text width=5ex,align=center},
		text depth=0.25ex,
		text height=1ex,
		nodes in empty cells
	},
	texto/.style={font=\footnotesize\sffamily},
	title/.style={font=\small\sffamily}
}
\begin{document}

\author{Belma Turkovic}
\affiliation{%
	\institution{Delft University of Technology}
	 	\city{Delft}
	 	\country{The Netherlands}
}
\email{B.Turkovic-2@tudelft.nl}
\author{Jorik Oostenbrink}
\affiliation{%
	\institution{Delft University of Technology}
	\city{Delft}
	\country{The Netherlands}
}
\email{J.Oostenbrink@tudelft.nl}
\author{Fernando Kuipers}
\affiliation{%
	\institution{Delft University of Technology}	
	 	\city{Delft}
	 	\country{The Netherlands}
}
\email{F.A.Kuipers@tudelft.nl}

\title{Detecting Heavy Hitters in the Data-plane}

\begin{abstract}
The ability to detect, in real-time, heavy hitters is beneficial to many network applications, such as DoS and anomaly detection. Through programmable languages as P4, heavy hitter detection can be implemented directly in the data-plane, allowing custom actions to be applied to packets as they are processed at a network node. This enables networks to immediately respond to changes in network traffic in the data-plane itself and allows for different QoS profiles for heavy hitter and non-heavy hitter traffic. 

Current interval-based methods that flush the whole counting structure are not well-suited for programmable hardware (the data-plane), because they either require more resources than available in that hardware, they do not provide good accuracy, or require too many actions from the control-plane. A sliding window approach that maintains accuracy over time would solve these issues. However, to the best of our knowledge, the concept of sliding windows in programmable hardware has not been studied yet.  

In this paper, we develop streaming approaches to detect heavy hitters in the data-plane. We consider the problems of (1) adopting a sliding window and (2) identifying heavy hitters separately and propose multiple memory- and processing-efficient solutions for each of them. These solutions are suitable for P4 programmable hardware and can be combined at will to solve the streaming variant of the heavy hitter detection problem.
\end{abstract}

\maketitle

\section{Introduction}\label{Sec_intro}
``Heavy hitter'' flows, i.e. flows with large traffic volumes, comprise less than 10\% of all flows in a data-center network, but carry most of the bytes transmitted in the network \cite{Benson2010}. Additionally, more than 80\% of flows last less than 11 seconds and carry less than than 10KB of data (just a few packets), while only $\approx 0.1\%$ last longer than 200s \cite{Benson2010, kandula2009nature}. This has interesting implications for traffic engineering, and quickly distinguishing between these two types of flows on a short time-scale is important for several applications such as DoS (Denial of Service) and anomaly detection, flow-size aware routing, and Quality of Service (QoS) management. 


Programmable switches, along with network programming languages such as P4  ~\cite{Bosshart2014}, offer new possibilities to detect heavy hitter flows directly in the data-plane while the packets are being processed. Consequently, specialized actions can be applied to these packets (e.g. providing higher or lower QoS or rerouting to avoid congestion), allowing network operators to respond to short traffic spikes quickly. This way, traffic flows belonging to applications that have very strict latency, jitter, and bandwidth requirements, such as the Tactile Internet, could be easily identified, enabling switches to treat them differently by providing per packet QoS \cite{belma}. 

Existing data-plane solutions such as HashPipe \cite{hashpipe} use memory and processing-efficient data-structures to count packets. However, they lack a mechanism to remove outdated information from the data-structure and rely on periodic flushing of the switch's memory. As a consequence, flows detected in the previous window are forgotten and need to be detected again each time the structure is flushed, thus decreasing accuracy and increasing detection time. In addition, flushing the memory of counting data-structures will lead to inconsistencies, as all memory can not be flushed simultaneously. This is especially prominent at switches that process hundreds of millions of packets every second.

A sliding window over the last $N$ packets solves the aforementioned problems by ensuring that only information about the last $N$ packets is present in the switch. This approach optimizes the detection time, increases accuracy, and has no need for special actions from the control-plane (e.g. register flushing) \cite{ben2016heavy}. However, despite these benefits, \emph{no efficient practical implementation of a heavy hitter algorithm using a sliding window targeting programmable networking devices exists}. 

Existing sliding window approaches use dynamic memory allocation or complex data-structures such as linked lists. Maintaining these structures requires many read/write actions, while switches with many 10-100GE ports have only a small time budget available if they want to maintain a high processing throughput (up to a few Tbps) \cite{hashpipe}. Existing hardware solutions that are optimized for low memory consumption (WCSS \cite{ben2016heavy}, Memento \cite{basat2018memento}) were not developed with P4 and programmable hardware in mind and generally exceed the available processing budget by using too many memory accesses per processed packet to maintain the window and counting structure.

In this paper, \emph{we present a solution for heavy hitter detection using a sliding window approach that is designed and optimized for programmable network hardware} by minimizing the processing overhead. That is, we minimize the additional number of cycles spent per packet to execute the heavy hitter algorithm. Additionally, in order to target different programmable hardware our solution is tunable with respect to memory usage and the number of stages in the switch. By increasing the available memory, the accuracy of our approach can be increased while keeping the processing time constant.


\section{Problem statement}
Detecting heavy hitters is a type of ``frequent items'' problem. That is, given a stream $S$, and a packet $p_y$ belonging to a flow $y$, the goal is to determine if more than $x\%$ of the last $N$ packets of $S$ belong to flow $y$. Often, an algorithm to solve this problem will do so by keeping track of frequency estimates $\widehat{f_y}$. 
Two possible errors can occur with such algorithms: (1) false positives, that is, falsely detecting a packet as belonging to a heavy hitter flow, and (2) false negatives, that is, failing to recognize a packet as belonging to a heavy hitter flow. While both of these errors should be minimized, false positives are typically preferred over false negatives, as accidentally ignoring heavy hitter traffic can have a significant impact. 

We identify two sub-problems: (1) keeping track of packet counts to determine if packets are heavy hitters (Sec. \ref{Sec_Sketches}), and (2) tracking the sliding window of $N$ packets by reducing the counts of packets that leave the window (Sec. \ref{Sec_Window}). We give multiple solutions for each of these sub-problems, which can be combined arbitrarily to solve the overall heavy hitter detection problem on programmable hardware.



\subsection{Hardware Constraints}
Switches, especially those at the core or those processing large amounts of data, have to process a large number of unique flows and only have limited hardware resources available. As a result, it is unfeasible and not scalable to store and maintain all flow frequencies in the data-plane. In general, compared to heavy hitter detection outside of the data-plane, the amount of memory available is much more limited, severely constraining any heavy hitter detection algorithm. 


More importantly, to avoid a drop in throughput, packets need to be processed as fast as they arrive (at line rate), only allowing for a processing budget of  nanoseconds. As an example, for a 100GE link and packets of size 64B the processing time per packet needs to be smaller than $6.88ns$. On current programmable hardware, memory accesses consume most processing cycles, so these should be limited as much as possible.  Typically, on some hardware just one read-modify-write action per each register array is allowed.


\section{Counting Sketch}\label{Sec_Sketches}
To keep track of frequency estimates and identify heavy hitters, we make use of sketches. Sketches are compact data structures that can be used to efficiently store large amounts of data. Instead of storing all data, they only store a summary of the data. This way, they trade in accuracy for memory. Typically, these data structures are probabilistic and make liberal use of hashing.


Sketches are usually optimized for low memory consumption, but do not track the flow identifiers of packets. However, as our goal is to identify heavy hitter packets while they are processed, storing flow identifiers is not needed.

\subsection{A hash table as the main building block}\label{Sec_hash}

Hash tables guarantee constant query and update time, have fixed memory footprint, and are supported by all programmable hardware. Thus, they are ideal as main building blocks for a counting sketch. 

As the amount of unique flows $k$ will often be significantly larger than the table width, there will be a large number of hash collisions. To keep the update and query time constant (and minimize the number of memory accesses), we do not resolve collisions (by exporting them to the CPU when a collision is detected as explained in \cite{p4measurments}), but simply seek to minimize their number and impact.


We define the load factor $\lambda_i$ of a hash table $i$ of size $width_i$ that stores flow statistics of $k_i$ unique flows entries as 
\begin{equation}
    \lambda_i = k_i/width_i
\end{equation}
This variable describes how filled up the table currently is. For example, a hash table with load factor $0.25$ is 25\% ``full.'' 

If the hash function is uniform, the number of flows $X_j$ that are mapped to a single table entry $j$ follows the binomial distribution $B(k_i,1/width_i)$. Now, the number of collisions of flow entry $j$, $C_j$, is \begin{equation}
    C_j = \begin{cases}0 & \text{if } X_j = 0\\X_j - 1 & \text{otherwise}\end{cases}
\end{equation}
Thus, the expected number of collisions of each table entry is
\begin{equation}
\begin{split}
    E[C_j] & = E[X_j-1] + P(X_j = 0) \\
    & = \lambda_i - 1 + (1 - 1/width_i)^{k_i}
 \end{split}
\end{equation}
As collisions directly scale with $k_i$, table widths should scale with the number of processed unique flows $k_i$ instead of the number of packets in the window $N$. 

The load factor $\lambda_i$ directly influences the probability of false positives. For every heavy hitter flow that is detected, on average, approximately $\lambda_i$ additional small flows are falsely identified. Additionally, multiple smaller flows can also either cause the heavy hitter flow to be detected prematurely or can add up together to the defined heavy hitter threshold. By adding more memory to our counting hash tables or by using multiple of them, a better accuracy can be achieved and the probability of false positives reduced (see Sec. \ref{Subsec_countmin} and Sec. \ref{Sec_Gated}). 


\subsection{Count-Min sketch}\label{Subsec_countmin}
As a first approach, we implement the Count-Min sketch \cite{countmin} in P4. The Count-Min sketch is a probabilistic data structure for storing frequencies. It consumes very little memory, but this comes at the cost of potentially overestimating frequencies. The Count-Min sketch stores frequencies in a two-dimensional array (multiple hash tables). 

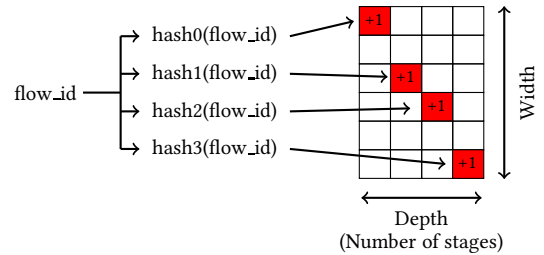
\begin{figure}[ht!]
    \begin{tikzpicture}

	\matrix[table] (mat11) 
	{
		|[fill=red]| {\footnotesize +1} & & & \\
		 & & & \\
		 & |[fill=red]| {\footnotesize +1} & & \\
		 & & |[fill=red]| {\footnotesize +1} & \\
		 & & & \\
		 & & & |[fill=red]| {\footnotesize +1} \\
	};
	\draw[<->, thick, color=black] (-0.8,-1.4) -- (0.8,-1.4);
	\node at (0,-1.7) {{\small Depth}};	
	\node at (0,-2.0) {{\small (Number of stages)}};	
	\draw[<->, thick, color=black] (1.1,1.15) -- (1.1,-1.15);
	\node[rotate=90] at (1.4,0) {{\small Width}};		
	\node at (-5,0) {{\small flow\_id}};	
	\node at (-2.75,0.75) {{\small hash0(flow\_id)}};	
	\node at (-2.75,0.25) {{\small hash1(flow\_id)}};	
	\node at (-2.75,-0.25) {{\small hash2(flow\_id)}};	
	\node at (-2.75,-0.75) {{\small hash3(flow\_id)}};	
	\draw[-, thick, color=black] (-4.5,0) -- (-4,0);
	\draw[-, thick, color=black] (-4,0.75) -- (-4,-0.75);	
	\draw[->, thick, color=black] (-4,0.75) -- (-3.75,0.75);
	\draw[->, thick, color=black] (-4,0.25) -- (-3.75,0.25);			
	\draw[->, thick, color=black] (-4,-0.25) -- (-3.75,-0.25);			
	\draw[->, thick, color=black] (-4,-0.75) -- (-3.75,-0.75);			
	
	\draw[->, thick, color=black] (-1.75,0.75) -- (-0.9,0.95);			
	\draw[->, thick, color=black] (-1.75,0.25) -- (-0.45,0.21);			
	\draw[->, thick, color=black] (-1.75,-0.25) -- (-0.15,-0.21);			
	\draw[->, thick, color=black] (-1.75,-0.75) -- (0.32,-0.95);					
					
\end{tikzpicture}
	\caption{Count-Min sketch.}\label{fig:countmin}
\end{figure}

The width of each table is smaller than the total number of unique flows, so flow identifiers are hashed to generate an index. To reduce the effect of collisions, the frequency of each flow is simultaneously maintained in multiple tables, each of which is indexed by a different hash function as shown in Fig.~\ref{fig:countmin}. The frequency is obtained by taking the minimum of these values. As a flow will collide with different flows in each hash table, the probability of falsely identifying a flow as a heavy hitter is reduced.

For a perfect window, if the width of the sketch is set to $\lceil \frac{e}{\epsilon} \rceil$, and the depth to $\lceil \frac{e}{\epsilon} \rceil$, the probability that the estimated frequency $\widehat{f_y}$ is smaller or equal to $ f_y + \epsilon \cdot N$ is at least $1 -\sigma$ \cite{countmin}. Thus, by increasing the width we can decrease the overestimation error bound $\epsilon$, while by increasing the depth we can increase the probability of staying within that error bound.

\textbf{P4 implementation.} When implementing Count-Min sketch in P4, one register array (to store the flow counts) and two match-action tables are needed per depth (except for $d=1$). The first match-action table updates the flow counts of the corresponding register array and makes sure that this count is always between $0$ and $N$. To calculate the index for this register array, we first calculate a hash of the packet identifier and then perform a modulo operation on that hash using the size of the register array. The second match-action table is used to determine the minimum between two flow counts from two successive register arrays. This way, as the packet passes through each table, the current minimum is always saved in a metadata variable that is at the end compared against the heavy hitter threshold. Finally, if the minimum count exceeds this heavy hitter threshold, a metadata variable to indicate this is set to 1 using a separate match-action table. 


\textbf{Memory consumption.} The total memory consumption of the data-structure presented in the previous subsection can be calculated as: 
\begin{equation}
    M_{count-min} = depth \cdot width \cdot \log_2(N)
\end{equation}




\subsection{Gated Sketch}\label{Sec_Gated}
\begin{figure}[ht!]
    	\begin{tikzpicture}
	
	\matrix[table] (mat11) at (-1.75,0)
	{
		\\
		|[fill=red]| {\tiny +1}  \\
		\\
		\\
		\\
		\\
		\\
		\\
		\\
		\\
		\\
		\\
	};
		\matrix[table] (mat11) at (-0.85,0)
	{
		\\
		\\
		|[fill=red]| {\tiny +1} \\
		\\
		\\
		 \\
		\\
		\\
	};
	
	\matrix[table] (mat11) at (-0.05,0)
	{
		\\
		\\
		\\
		|[fill=red]| {\tiny +1}  \\
		\\
		\\
	};
	\matrix[table] (mat11) at (0.7,0)
	{
		\\
		\\
		\\
		|[fill=red]| {\tiny +1}  \\
	};
	%
	\draw[<->, thick, color=black] (-2,2.7) -- (1,2.7);
	\node at (-0.5,2.5) {{\small Depth = 4}};	
	\node at (-0.5,2.9) {{\small (Number of stages)}};	
	\draw[<->, thick, color=black] (1.1,0.77) -- (1.1,-0.77);
	\node[rotate=90] at (1.4,0) {{\small Width\_3}};		
	\node at (-6.5,1.35) {{\small id}};	
	\node at (-3.15,0.75) {{\small hash0(id)}};
	\node at (-4.2, 0.25) {{\small if(hash0(id) > th0) hash1(id)}};	
	\node at (-4.2,-0.25) {{\small if(hash1(id) > th1) hash2(id)}};	
	\node at (-4.2,-0.75) {{\small if(hash2(id) > th2) hash3(id)}};	
	\draw[-, thick, color=black] (-6.5,0) -- (-6.2,0);
	\draw[-, thick, color=black] (-6.2,0.75) -- (-6.2,-0.75);	
	\draw[->, thick, color=black] (-6.2,0.75) -- (-3.75,0.75);
	\draw[->, thick, color=black] (-6.2,0.25) -- (-5.85,0.25);			
	\draw[->, thick, color=black] (-6.2,-0.25) -- (-5.85,-0.25);			
	\draw[->, thick, color=black] (-6.2,-0.75) -- (-5.85,-0.75);			
	\draw[-, thick, color=black]  (-6.5,1.15) -- (-6.5,0);
			
	\draw[->, thick, color=black] (-2.55,0.75) -- (-2.05,1.8);			
	\draw[->, thick, color=black] (-2.55,0.25) -- (-1.15,0.61);			
	\draw[->, thick, color=black] (-2.55,-0.25) -- (-0.4,-0.19);			
	\draw[->, thick, color=black] (-2.55,-0.75) -- (0.37,-0.55);					

	\end{tikzpicture}
	\caption{Gated sketch.}\label{fig:gated}
\end{figure}
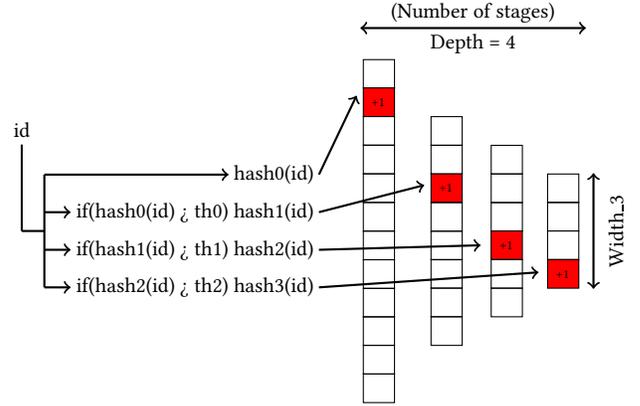

To use the switch's memory more efficiently, we have developed a new sketch. It uses a set of hash tables of different widths, but, unlike the Count-Min sketch, does not update every hash table for each processed packet. When a new packet arrives, counters from the hash tables are compared against a set of thresholds $th_0$ to $th_{d-1}$ whose sum equals the heavy hitter threshold $th$: 
\begin{equation}
    th = \sum_{i=0}^{d-1} {th_i}
\end{equation}
where $th_i$ is the threshold of hash table $i$. The packet is only processed by the next table $i+1$ if the counter value of the current table $i$ is higher than its threshold $th_i$ (see Fig. \ref{fig:gated}). If the counters from all hash tables satisfy their respective thresholds, the packet is identified as a heavy hitter.

This approach has multiple advantages. First, the number of collisions at deeper hash tables is reduced, as less packets are processed by them. As a consequence, the width of the deeper tables can be reduced without losing much accuracy. This reduces the overall memory consumption and makes it possible to trade-off the width of the deeper tables for the width of the first table. 

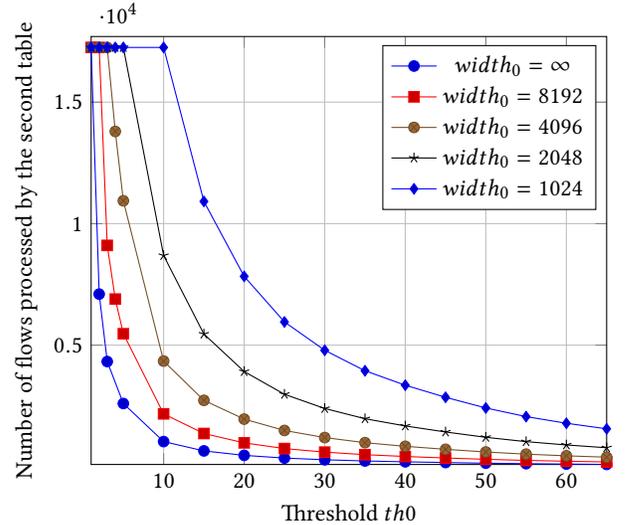
\begin{figure}[ht!]
	\begin{tikzpicture}
		\begin{axis}[
		    ylabel={Number of flows processed by the second table},
		    xlabel={Threshold $th0$},
		    xmin=1, xmax=65, ymin=90, ymax=17700, grid,
		    ]
		    
		    \addplot+ []
    		table [x index=0, y index=1, col sep=comma] 
            {
		    1,  17250.63
            2,  7101.39
            3,  4325.33
            5,  2598.47
            10, 1031.80
            15, 648.05
            20, 464.87
            25, 353.62
            30, 284.36
            35, 234.31
            40, 198.98
            45, 169.47
            50, 143.40
            55, 122.29
            60, 106.05
            65, 92.420
		    };
		    
            \addplot+ []
    		table [x index=0, y index=1, col sep=comma] 
            {
		    1,  17250.63
		    2,  17250.63
			3,  9105.958 
			4,  6896.484 
			5,  5470.463 
			10, 2172.211 
			15, 1364.316 
			20,  978.674 
			25,  744.463 
			30,  598.653 
			35,  493.284 
			40,  418.905 
			45,  356.779 
			50,  301.895 
			55,  257.453 
			60,  223.263 
			65,  194.568 
			100,  92.632 
            };
            
            \addplot+ []
    		table [x index=0, y index=1, col sep=comma] 
            {
		    1, 17250.63
		    2, 17250.63
		    3, 17250.63
			4, 13792.968
			5, 10940.926
			10, 4344.421
			15, 2728.632
			20, 1957.347
			25, 1488.926
			30, 1197.305
			35,  986.568
			40,  837.811
			45,  713.558
			50,  603.789
			55,  514.905
			60,  446.526
			65,  389.137
			100, 185.263
            };            
             \addplot+ []
    		table [x index=0, y index=1, col sep=comma] 
            {
		    1,  17250.63
		    2,  17250.63
		    3,  17250.63
		    4,  17250.63
		    5,  17250.63
			10, 8688.842
			15, 5457.263
			20, 3914.695
			25, 2977.853
			30, 2394.611
			35, 1973.137
			40, 1675.621
			45, 1427.116
			50, 1207.579
			55, 1029.811
			60,  893.053
			65,  778.274
			100, 370.526
            };     
             \addplot+ []
    		table [x index=0, y index=1, col sep=comma] 
            {
		    1,  17250.63
		    2,  17250.63
		    3,  17250.63
		    4,  17250.63
		    5,  17250.63
		    10,  17250.63
			15, 10914.526
  		    20,  7829.389
  		    25,  5955.705
  		    30,  4789.221
  		    35,  3946.274
  		    40,  3351.242
  		    45,  2854.232
  		    50,  2415.158
  		    55,  2059.621
  		    60,  1786.105
  		    65,  1556.547
  		    100,  741.053
            };
           
            \legend{$width_0=\infty$, $width_0=8192$, $width_0=4096$,  $width_0=2048$, $width_0=1024$  }
		\end{axis}
	\end{tikzpicture}
	\caption{Average number of flows processed by the second stage depending on the width of the first stage ($width_0$) and the threshold of the first stage ($th_0$). Calculated using CAIDA traces \cite{caida2016trace} from an ISP backbone link.}\label{fig:flowstats}
\end{figure}

Fig. \ref{fig:flowstats} shows the average number of flows processed by the second stage depending on the width of the first stage ($width_0$) and the threshold of the first stage ($th_0$) calculated using CAIDA traces from 2016 collected on an ISP backbone router \cite{caida2016trace}. If we choose $th_0=0$, all packets are processed by the deeper stages (as in the Count-Min sketch). By increasing the value of $th_0$ the number of unique flows processed by the second stage drops significantly. Similarly, by increasing the value of $width_0$, the number of flows that pass to the second stage due to collisions decreases.


Second, by using a Gated Sketch the average processing time per packet can be reduced, as many packets are not processed at deeper tables.

Finally, the P4 implementation is simpler and uses the switch resources much more efficiently than the Count-Min sketch, as explained bellow. 
  

\textbf{P4 implementation.} When implementing the gated sketch in P4 only one register array and one match-action table, to maintain the flow counts, are needed per depth. The match-action table is used to update the flow count in the corresponding register array. Additionally, before a match-action table is applied in the ingress control block, it is checked if the count from the previous table satisfies its respective threshold. This significantly simplifies the design when compared to the Count-Min sketch and the number of tables that are needed is reduced by a factor of 2. 

\textbf{Memory consumption.} The total memory consumption of the data-structure presented in the previous subsection can be calculated as: 
\begin{equation}
    M_{gated} = \sum_{i = 0}^{d-1}{width_{i} \cdot \log_2{(N-\sum_{i=0}^{d-i-1}{th_i})}}
\end{equation}



\section{Sliding window}\label{Sec_Window}
The sketching approaches described in the previous section count all packets that are received by the node since it was started (or since the register values were reset). However, only recent packets in the stream are relevant and represent the current state in the network. 

If register values are reset every $N$ packets, the probability of false negatives at the beginning of the window can be significant. Additionally, resetting all the counts on a switch requires an action from the control-plane. In case of a heavy hitter sketch that is running in the data-plane, the state in the switches (e.g. flow counts) will change at line rate (at speeds that can reach Tbps), preventing any software-based controller from consistently resetting all the used register arrays. Additionally, small window sizes, such as $2^{16}$ packets, correspond to not more than a fraction of a second on a 100Gbps link. Resetting a state from a controller on such a short time-scale is ineffective, requires too many actions and is possible only based on time (every $T$ seconds), and not on the number of packets.

This problem can be solved with a sliding window over the last $N$ packets. If such a data structure is added to the counting sketch, outdated flows and counts can be removed from the counting sketch. 

\subsection{Ring sliding window}\label{SubSec_ringWindow}
\begin{figure}[ht!]
	\begin{tikzpicture}

		\matrix[tablemiddle] (mat11) at (0,0)
	{
		\\
		|[fill=tudelft]| {\scriptsize -1}  \\
		\\
			\\	\\	\\	\\	\\	\\
	};
	
	\draw[<->, thick, color=black] (0.6,1.75) -- (0.6,-1.75);
	\node[rotate=90] at (0.8,0) {{\small Width\_0}};	
		
	\node at (-1.95,0) {{\small h0(flow\_id3)}};	
	\draw[->, thick, color=black] (-3.5,0) -- (-2.75,0);			
	
	\draw[->, thick, color=black] (-1.15,0) -- (-0.35,1);			
	
	\matrix[tablelong]  at (-4,0) (mat11) 
	{
		|[fill=tudelft]| {\footnotesize flowid0} \\
		|[fill=tudelft]| {\footnotesize flowid1} \\
		|[fill=tudelft]| {\footnotesize flowid2} \\
		|[fill=tudelft]| {\footnotesize flowid3} \\
		|[fill=tudelft]| {\footnotesize flowid4} \\
		|[fill=tudelft]| {\footnotesize flowid5} \\
		|[fill=tudelft]| {\footnotesize flowid6} \\
	};
	\node at (-5.5,0) {{\small index}};	
	\draw[->, thick, color=black] (-5.1,0) -- (-4.6,0);					
\end{tikzpicture}
	\caption{Ring implementation of the sliding window. Reducing the counts in the tables.}\label{fig:ring}
\end{figure}
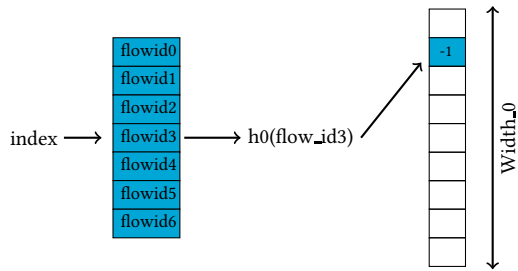

Our first approach in implementing a sliding window is to keep track of the flow identifiers for the last $N$ packets in an array, similarly to the way described in \cite{assaf2017,basat2017}. Every time a new packet arrives the oldest entry from the array is removed and replaced with the new flow identifier. Afterwards, counts for the flow that was removed are reduced in all hash tables as shown in Fig. \ref{fig:ring}.

The main advantage of this approach is high accuracy. All hash tables only contain the counts of the last $N$ packets, and the probability of false negatives is equal to 0. In case of collisions, the frequency of flows can be overestimated, but can never be underestimated. However, as the ring structure takes up a large amount of memory, it is not practical for larger window sizes. The array of flow identifiers (5-tuples) shown in Fig. \ref{fig:ring} takes up $13 \times N$ bytes and the index register, used to store the position of the oldest packet in the ring, an additional $log_2(N)$ bytes.

\textbf{Memory consumption.}  To save memory, it is possible to just store the values $h_i(flow\_id)$ for each depth. Thus, the memory consumption of the presented structure is equal to: 
\begin{equation}
    M_{ring} = N \cdot \sum_{i = 0}^{d-1}{log_2(width_{i})}
\end{equation}

\begin{figure}[ht!]
    	\begin{tikzpicture}[scale = 0.85]
		\begin{semilogxaxis}[
		    ylabel={Memory consumption $\lbrack MB \rbrack$},
		    xlabel={$N$},
		    xmin=32767, xmax=2097152, ymin=0, ymax=2, grid, log basis x=2,
		    legend style={at={(0.61,0.01)},anchor=south west} 
		    ]
            \addplot+ [mark=none, domain=1:2097152, red, ultra thick ]{1.4};
            \addplot+ [blue,mark options={fill=blue}]
    		table [x index=0, y index=1, col sep=comma] 
            {
                 32768, 0.049152
                 65536, 0.098304
                 131072, 0.196608
                 262144, 0.393216
                 524288, 0.786432
                 1048576, 1.572864
                 2097152, 3.145728
		    };
            \addplot+ [magenta,mark options={fill=magenta}]
    		table [x index=0, y index=1, col sep=comma] 
            {		    
                 32768, 0.098304
                 65536, 0.196608
                 131072, 0.393216
                 262144, 0.786432
                 524288, 1.572864
                 1048576, 3.145728
                 2097152, 6.291456
            };
            
            \addplot+ []
    		table [x index=0, y index=1, col sep=comma] 
            {		    
                 32768, 0.147456
                 65536, 0.294912
                 131072, 0.589824
                 262144, 1.179648
                 524288, 2.359296
                 1048576, 4.718592
                 2097152, 9.437184
                };        
            \addplot+ []
    		table [x index=0, y index=1, col sep=comma] 
            {		    
                32768, 0.425984
                 65536, 0.851968
                 131072, 1.703936
                 262144, 3.407872
                 524288, 6.815744
                 1048576, 13.631488
                 2097152, 27.262976
            };
                
		   \legend{\footnotesize Hardware Limit, \footnotesize $depth =1$, \footnotesize $depth=2$, \footnotesize $depth=3$, \footnotesize $flow\_id \; (13B)$}
        \end{semilogxaxis}
	\end{tikzpicture}
	\caption{Memory consumption of the ring window used together with the Count-Min sketch with $width = 4096$.}\label{fig:seqmem}
\end{figure}
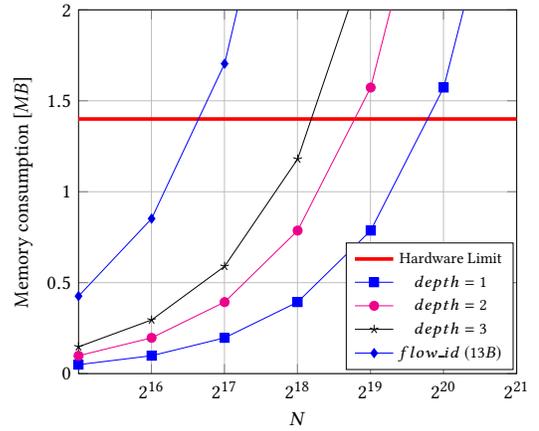

Since programmable switches have limited memory (typically 1.4MB per stage \cite{hashpipe}), the ring structure becomes infeasible for $N \geq 2^{20}$ (between $ 0.1$ and $1.3$ s on a 10Gbps link) even if the depth of the counting structure is equal to 1 (Fig. \ref{fig:seqmem}). By increasing the depth of the counting structure, the memory doubles.

\textbf{P4 Implementation.} 
To implement this structure in P4, we need to either save $N$ packet identifiers (5-tuple) or all the indices for counting register arrays that were increased while the last $N$ packets were processed. In the first case, 5 additional register arrays are needed to store the flow identifier (source and destination IP address, protocol field and source and destination port). In the second case, $d$ additional register arrays are needed to store the indices for each counting register array. Before a new packet is processed, an additional table to read the flow identifier (or all the register indices) of the oldest packet is applied. This significantly increases the overhead of the heavy hitter algorithm, since the number of memory accesses per register array (typically one read-modify-write action is available) as well as the total number of register accesses is limited on most programmable switches and can lead to drops in throughput. 

For every incoming packet, two counts need to be modified for each counting register array: (1) the count of the flow of the newest packet is increased and (2) the count of the flow of the oldest packet in the window is decreased. As a consequence, this solution is not feasible on programmable hardware that stores the register values in the local memory and has a limit on the number of read-modify-write actions per register. On programmable hardware that uses shared memory (e.g. Netronome) this limitation is not present, but the number of memory accesses is high and can cause a drop in throughput. In addition, as packets are processed in parallel, shared memory can lead to race conditions. As a consequence, the probability of false positives as well as false negatives (which was 0) will increase. 

\subsection{Sequential sliding window}\label{SubSec_seqWindow}

To develop a solution that is feasible on programmable hardware as well as for larger values of $N$ (in contrast to the previously described Ring window), we have developed a solution that only needs $log_2(width)$ additional bits to maintain the sliding window. 
Every time a packet is added to the sketch, we also reduce all counts in a row (determined by a sequential index) as shown in Fig. \ref{fig:sequential}.

\begin{figure}[ht!]
    \begin{tikzpicture}
	\node at (0,1.3) {{\small pkt0}};	
	\node at (3.85,1.3) {{\small pkt1}};	
	\matrix[table] (first) 
	{
		|[fill=tudelft]| {\footnotesize -1} & |[fill=tudelft]| {\footnotesize -1} & |[fill=tudelft]| {\footnotesize -1} & |[fill=tudelft]| {\footnotesize -1} \\
		& & &  \\
		& & &  \\
		& & &  \\
		& & &  \\
		& & &  \\
	};
	\node at (-1.5,1.2) {{\small index = 0}};	
	\draw[->, thick, color=black] (-1.6,0.95) -- (-1.1,0.95);					
	\draw[<->, thick, color=black] (-0.8,-1.4) -- (0.8,-1.4);
	\node at (0,-1.7) {{\small Depth}};	
	\node at (0,-2.0) {{\small (Number of stages)}};	
	\draw[<->, thick, color=black] (1.1,1.15) -- (1.1,-1.15);
	\node[rotate=90] at (1.4,0) {{\small Width}};		
	\matrix[table] (second) at (4,0)
    {
		& & &  \\
		|[fill=tudelft]| {\footnotesize -1} & |[fill=tudelft]| {\footnotesize -1} & |[fill=tudelft]| {\footnotesize -1} & |[fill=tudelft]| {\footnotesize -1} \\
		& & &  \\
		& & &  \\
		& & &  \\
		& & &  \\
    };
	\node at (2.5,0.85) {{\small index = 1}};	
	\draw[->, thick, color=black] (2.4,0.6) -- (2.9,0.6);
	\draw[<->, thick, color=black] (3.2,-1.4) -- (4.8,-1.4);
	\node at (4,-1.7) {{\small Depth}};	
	\node at (4,-2.0) {{\small (Number of stages)}};	
	\draw[<->, thick, color=black] (5.1,1.15) -- (5.1,-1.15);
	\node[rotate=90] at (5.4,0) {{\small Width}};	


\end{tikzpicture}
	\caption{Sequential implementation of the sliding window. Reducing the counts in the tables.}\label{fig:sequential}
\end{figure}

In this scheme, the probability of false positives and false negatives can be significant, as, in contrast to the ring implementation, we do not reduce the flow counts of the oldest packet of the sliding window. Moreover, many entries in the tables can be 0 (depending on the width of the table). As these counts can not be further reduced, and one other count will be increased, the total number of counts increased per window can be larger than the total number of counts reduced. As these counts are never removed, accuracy decreases over time. Additionally, when a heavy hitter flow is completed, it takes many cycles for that flow to be removed from the tables causing potential collisions with the newer flows and increasing, as a consequence, the probability of false positives. False negatives are possible since, at the time an entry is increased in the table, the same entry can be removed. 

However, the simplicity and the fact that only $log_2(width)$ additional bits of memory are needed makes this approach suitable for programmable network hardware. 

\textbf{P4 Implementation.} When implementing this scheme in P4, just one additional index needs to be maintained. Every time a match-action table is applied to update the count for the newest packet, one count from the same register array is reduced using this sequential index. Afterwards, the sequential index is increased by 1 and saved.

In addition, this scheme is easily implementable on all programmable hardware, as the number of memory accesses per each register array can be reduced to one. To do this the counting register array needs to be split in two tables, as shown in Fig. \ref{fig:sequential2}. 
\begin{figure}[ht!]
    \begin{tikzpicture}
	\matrix[table] (first) 
	{
		|[fill=tudelft]| {\footnotesize -1} \\
		  \\
		  \\
		  		  \\
	};
	\node at (-1.1,0.7) {{\small index0\_0 = 0}};
	\draw[->, thick, color=black] (-1.4,0.55) -- (-0.5,0.55);					
	\node at (-1,-1.35) {{\small hash0(id)}};	
	\draw[->, thick, color=black] (-1.4,-1.55) -- (-0.5,-1.55);					

	\matrix[table] (first1) at (0, -1.75)
	{
		  \\
	      |[fill=red]| {\footnotesize +1} \\
		  \\
		  		  \\
	};
	\node at (0,-2.75) {{\small Stage 0}};	

	\node at (2.9,-0.95) {{\small index1\_1 = 0}};	
	\draw[->, thick, color=black] (2.6,-1.15) -- (3.5,-1.15);					
	\node at (3,0.7) {{\small hash1(id)}};
	\draw[->, thick, color=black] (2.6,0.55) -- (3.5,0.55);					

	\matrix[table] (second) at (4,0)
	{
		|[fill=red]| {\footnotesize +1} \\
		  \\
		  \\
		  		  \\
	};
	
	\matrix[table] (second2) at (4, -1.75)
	{
	      |[fill=tudelft]| {\footnotesize -1} \\
		  \\
		  		  \\
		  		  		  \\
	};
	\node at (4,-2.75) {{\small Stage 1}};

	\draw[<->, thick, color=black] (-0.4,-3.1) -- (4.5,-3.1);
	\node at (2,-3.4) {{\small Depth = 2}};	
	\node at (2,-3.7) {{\small (Number of stages)}};	
	\draw[<->, thick, color=black] (1.3,0.75) -- (1.3,-2.5);
	\node[rotate=90] at (1.6,-0.85) {{\small Width}};		
	\node[rotate=90] at (1,0) {{\small Width/2}};
	\draw[<->, thick, color=black] (0.8,0.75) -- (0.8,-0.75);
	\draw[<->, thick, color=black] (0.8,-1) -- (0.8,-2.5);

	\node[rotate=90] at (1,-1.75) {{\small Width/2}};		


\end{tikzpicture}
	\caption{Sequential implementation of the sliding window. Reducing the counts in the tables.}\label{fig:sequential2}
\end{figure}
Depending on the value of the $hash(id)$ (hash of the flow identifier of the first received packet), a value from either the first or the second table will be decreased. As a consequence, the total memory consumption of this extended data-structure is increased and equal to:  
\begin{equation}
    M_{seq} =  2\cdot \sum_{i = 0}^{d-1}{log_2(width_{i})} 
\end{equation}
as 2 indices (one for each half of the table) are needed per depth.

\subsection{Sequential flushing}

\begin{figure}[ht!]
    \begin{tikzpicture}
	\node at (0,1.3) {{\small pkt\_0}};	
	\node at (3.85,1.3) {{\small pkt\_m}};	
	\matrix[table] (first) 
	{
		|[fill=tudelft]| {\footnotesize 0} & |[fill=tudelft]| {\footnotesize 0} & |[fill=tudelft]| {\footnotesize 0} & |[fill=tudelft]| {\footnotesize 0} \\
		& & &  \\
		& & &  \\
		& & &  \\
		& & &  \\
		& & &  \\
	};
	\node at (-1.5,1.2) {{\small index = 0}};	
	\draw[->, thick, color=black] (-1.6,0.95) -- (-1.1,0.95);					
	\draw[<->, thick, color=black] (-0.8,-1.4) -- (0.8,-1.4);
	\node at (0,-1.7) {{\small Depth}};	
	\node at (0,-2.0) {{\small (Number of stages)}};	
	\draw[<->, thick, color=black] (1.1,1.15) -- (1.1,-1.15);
	\node[rotate=90] at (1.4,0) {{\small Width}};		
	\matrix[table] (second) at (4,0)
    {
		& & &  \\
		|[fill=tudelft]| {\footnotesize 0} & |[fill=tudelft]| {\footnotesize 0} & |[fill=tudelft]| {\footnotesize 0} & |[fill=tudelft]| {\footnotesize 0} \\
		& & &  \\
		& & &  \\
		& & &  \\
		& & &  \\
    };
	\node at (2.5,0.85) {{\small index = 1}};	
	\draw[->, thick, color=black] (2.4,0.6) -- (2.9,0.6);
	\draw[<->, thick, color=black] (3.2,-1.4) -- (4.8,-1.4);
	\node at (4,-1.7) {{\small Depth}};	
	\node at (4,-2.0) {{\small (Number of stages)}};	
	\draw[<->, thick, color=black] (5.1,1.15) -- (5.1,-1.15);
	\node[rotate=90] at (5.4,0) {{\small Width}};	


\end{tikzpicture}
	\caption{Sequential flushing.}\label{fig:sequentialf}
	\vspace{-0.3cm}
\end{figure}
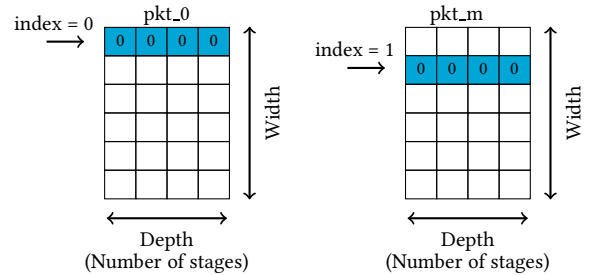

The main idea of this approach is to reset the counting structure in every window $N$. For every $m$-th (where $m=N/width$) packet that is added to the sketch, we also reduce all counts in a row (determined by a sequential  index) as shown in Fig. \ref{fig:sequentialf}. This way, after $N$ (window) packets are processed, all the registers values have been reset to 0. The main advantage of this scheme, in contrast to the Sequential window, is that it can maintain accuracy over time since the whole structure is reset every $N$ packets.

In this scheme,  false positives and false negatives will always be present, as in contrast to the ring implementation we do not reduce the flow counts of the oldest packet of the sliding window. In addition, counts across columns are inconsistent, as we flush the counts of different flows in each column. 

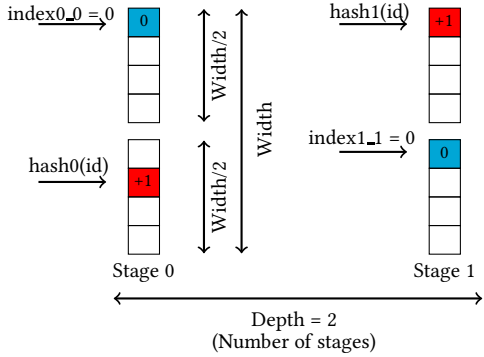
\begin{figure}[ht!]
    \begin{tikzpicture}
	\matrix[table] (first) 
	{
		|[fill=tudelft]| {\footnotesize 0} \\
		  \\
		  \\
		  		  \\
	};
	\node at (-1.1,0.7) {{\small index0\_0 = 0}};
	\draw[->, thick, color=black] (-1.4,0.55) -- (-0.5,0.55);					
	\node at (-1,-1.35) {{\small hash0(id)}};	
	\draw[->, thick, color=black] (-1.4,-1.55) -- (-0.5,-1.55);					

	\matrix[table] (first1) at (0, -1.75)
	{
		  \\
	      |[fill=red]| {\footnotesize +1} \\
		  \\
		  		  \\
	};
	\node at (0,-2.75) {{\small Stage 0}};	

	\node at (2.9,-0.95) {{\small index1\_1 = 0}};	
	\draw[->, thick, color=black] (2.6,-1.15) -- (3.5,-1.15);					
	\node at (3,0.7) {{\small hash1(id)}};
	\draw[->, thick, color=black] (2.6,0.55) -- (3.5,0.55);					

	\matrix[table] (second) at (4,0)
	{
		|[fill=red]| {\footnotesize +1} \\
		  \\
		  \\
		  		  \\
	};
	
	\matrix[table] (second2) at (4, -1.75)
	{
	      |[fill=tudelft]| {\footnotesize 0} \\
		  \\
		  		  \\
		  		  		  \\
	};
	\node at (4,-2.75) {{\small Stage 1}};

	\draw[<->, thick, color=black] (-0.4,-3.1) -- (4.5,-3.1);
	\node at (2,-3.4) {{\small Depth = 2}};	
	\node at (2,-3.7) {{\small (Number of stages)}};	
	\draw[<->, thick, color=black] (1.3,0.75) -- (1.3,-2.5);
	\node[rotate=90] at (1.6,-0.85) {{\small Width}};		
	\node[rotate=90] at (1,0) {{\small Width/2}};
	\draw[<->, thick, color=black] (0.8,0.75) -- (0.8,-0.75);
	\draw[<->, thick, color=black] (0.8,-1) -- (0.8,-2.5);

	\node[rotate=90] at (1,-1.75) {{\small Width/2}};		


\end{tikzpicture}
	\caption{Sequential flushing.}\label{fig:sequentialf2}
	\vspace{-0.45cm}
\end{figure}

Similarly to the sequential window, the simplicity and the fact that this solution only needs additional $depth\cdot log_2(width)$ bits of memory makes this approach suitable for programmable network hardware. Just as the sequential window, this solution is implementable on all available P4 hardware using the extension presented in Fig. \ref{fig:sequentialf2}.

\subsection{Hybrid window}\label{Sliding_window_impl}

This approach improves upon the previously implemented ring structure (Sec. \ref{SubSec_ringWindow}). The main disadvantage of the ring approach is its high memory consumption: every flow identifier of the last $N$ packets needs to be stored inside a register array of size $N$. 

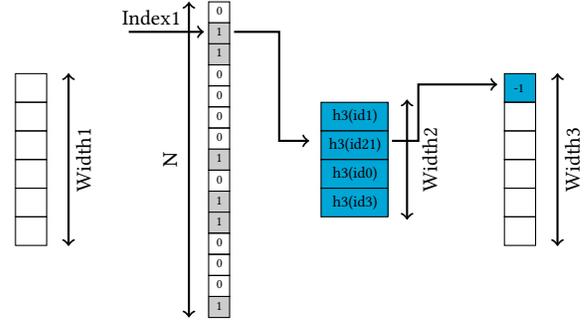
\begin{figure}[ht!]
	\begin{tikzpicture}

	
		\matrix[table] (mat11) at (-6,0)
	{
		\\
		\\
		\\
		\\
		\\ \\
	};

	\matrix[table] (mat11) at (0.5,0)
	{
		|[fill=tudelft]| {\tiny -1} \\
		\\
		\\
		\\
		\\ \\
	};

	
	

	\node[rotate=90] at (-4.15,0) {{\small N}};		
	\draw[<->, thick, color=black] (-3.9,-2.1) -- (-3.9,2.1);

	\matrix[tablesmall] (mat11) at (-3.5,0)
	{
		{\tiny 0} \\ |[fill=gray!40]| {\tiny 1} \\ |[fill=gray!40]| {\tiny 1} \\ {\tiny 0}
		\\ {\tiny 0} \\  {\tiny 0} \\ {\tiny 0}
		\\ |[fill=gray!40]| {\tiny 1} \\  {\tiny 0} \\ |[fill=gray!40]| {\tiny 1}
		\\ |[fill=gray!40]| {\tiny 1} \\  {\tiny 0} \\ 	{\tiny 0}	
		\\  {\tiny 0} \\ |[fill=gray!40]| {\tiny 1}  \\
	};
	
	\matrix[tablelong] (mat11) at (-1.7,0)
	{
		|[fill=tudelft]| {\scriptsize h3(id1)} \\ |[fill=tudelft]| {\scriptsize h3(id21)} \\ 
		|[fill=tudelft]| {\scriptsize h3(id0)} \\ |[fill=tudelft]| {\scriptsize h3(id3)} \\ 
	};

	\draw[->, thick, color=black] (-3.3,1.7) -- (-2.7,1.7)  -- (-2.7,0.25) -- (-2.3,0.25);
	\draw[->, thick, color=black] (-1.2,0.25) -- (-0.85,0.25)  -- (-0.85,1) -- (0.2,1);
	\draw[->, thick, color=black] (-4.7,1.7) -- (-3.7,1.7);
	\node[] at (-4.4,1.9) {\small Index1};	
	


	\draw[<->, thick, color=black] (-5.5,1.15) -- (-5.5,-1.15);
	\node[rotate=90] at (-5.3,0) {{\small Width1}};	
	\draw[<->, thick, color=black] (1,1.15) -- (1,-1.15);
	\node[rotate=90] at (1.2,0) {{\small Width3}};	
	\draw[<->, thick, color=black] (-1,0.8) -- (-1,-0.8);
	\node[rotate=90] at (-0.7,0) {{\small Width2}};

\end{tikzpicture}
	\caption{Hybrid implementation of the sliding window.}\label{fig:hybrid2}
		\vspace{-0.3cm}
\end{figure}

To reduce memory usage, we propose a new ring structure that stores a smaller number of identifiers (Fig. \ref{fig:hybrid2}). Instead of removing packets from the counting sketch as soon as they leave the window, our structure removes packets in batches of $th\cdot N /m ,$ (threshold as a percentage times $N/m$) at a time (similarly to \cite{ben2016heavy}). To keep track of heavy hitters, it adds an additional structure -- for counting the number of times entries reached $th\cdot N / m$ -- to the sketch (shown on the right side of the Fig. \ref{fig:hybrid2}). Every time an entry of the sketch reaches $th\cdot N/m$, the entry is set to 0, and we increase the count of the right structure by 1. Now, to identify if a packet is a heavy hitter, we check if this count is larger or equal than $m$. To remove a batch of $th \cdot N/m$ packets, we simply reduce the count by 1. To make sure our window is of size $N$, we reduce the $th\cdot N/m$ count of an entry exactly $N$ packets after we increase it.  

This window structure is implemented using two arrays: (1) a flowid array containing flow identifiers of packets that reached $th\cdot N/m$ (third array in Fig. \ref{fig:hybrid2}) and (2) a bit array of size $N$ specifying when the $th \cdot N / m$ count was increased (second array in Fig. \ref{fig:hybrid2}). In addition to the counting sketch itself (to count up to $th\cdot N/m$), this approach requires an additional counting sketch to count the number of times every count reached $th\cdot N/m$.  

In order to implement the two data structures needed to maintain the window (flowid array and the bit array) three additional indices are needed: (1) \textbf{index1} used to keep track of the current position in the window of size $N$, (2) \textbf{first} used to keep track of the place in the flowid at which a new packet will be added, and (3) \textbf{last} used to point to the location in flowid that is storing the oldest entry that was added. 

If a batch of $th \cdot N / m$ packets needs to be removed (the value of the bit array is 1), a value from flowid is read using the \textbf{last} index. Subsequently, that flowid row is set to 0, and the value of \textbf{last} incremented by 1 to point to the new oldest item as shown in Fig. \ref{fig:hybridadd}. Similarly, if a packet is added to the flowid array, the value of \textbf{first} is incremented by 1 and the value of the flowid row updated (Fig. \ref{fig:hybridadd2}). 

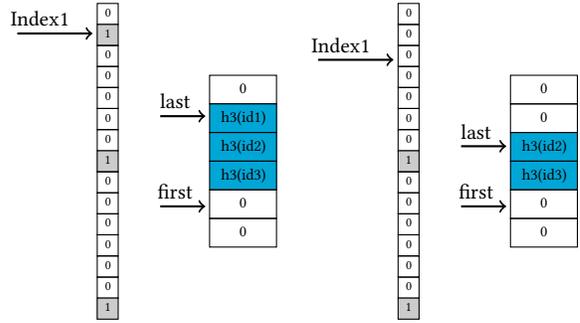
\begin{figure}[ht!]
	\begin{tikzpicture}
	
	\matrix[tablelong] (mat11) at (-1.7,0)
	{
		{\scriptsize 0} \\ |[fill=tudelft]| {\scriptsize h3(id1)} \\ |[fill=tudelft]| {\scriptsize h3(id2)} \\ 
		|[fill=tudelft]| {\scriptsize h3(id3)} \\ \scriptsize 0 \\ {\scriptsize 0} \\
	};
	\matrix[tablelong] (mat11) at (2.3,0)
	{
		{\scriptsize 0} \\ {\scriptsize 0} \\ |[fill=tudelft]| {\scriptsize h3(id2)} \\ 
		|[fill=tudelft]| {\scriptsize h3(id3)} \\ \scriptsize 0 \\ {\scriptsize 0} \\
	};	
	
	\matrix[tablesmall] (mat11) at (-3.5,0)
	{
		{\tiny 0} \\ |[fill=gray!40]| {\tiny 1} \\ {\tiny 0} \\ {\tiny 0}
		\\ {\tiny 0} \\  {\tiny 0} \\ {\tiny 0}
		\\ |[fill=gray!40]| {\tiny 1} \\  {\tiny 0} \\ {\tiny 0}
		\\ {\tiny 0} \\  {\tiny 0} \\ 	{\tiny 0}	
		\\  {\tiny 0} \\ |[fill=gray!40]| {\tiny 1}  \\
	};

	\matrix[tablesmall] (mat11) at (0.5,0)
	{
		{\tiny 0} \\ {\tiny 0} \\ {\tiny 0} \\ {\tiny 0}
		\\ {\tiny 0} \\  {\tiny 0} \\ {\tiny 0}
		\\ |[fill=gray!40]| {\tiny 1} \\  {\tiny 0} \\ {\tiny 0}
		\\ {\tiny 0} \\  {\tiny 0} \\ 	{\tiny 0}	
		\\  {\tiny 0} \\ |[fill=gray!40]| {\tiny 1}  \\
	};	
	
	\draw[->, thick, color=black] (-4.7,1.7) -- (-3.7,1.7);
	\node[] at (-4.4,1.9) {\small Index1};	
	\draw[->, thick, color=black] (-2.8,0.6) -- (-2.2,0.6);
	\node[] at (-2.6,0.8) {\small last};	
    \draw[->, thick, color=black] (-2.8,-0.6) -- (-2.2,-0.6);
	\node[] at (-2.6,-0.4) {\small first};

	\draw[->, thick, color=black] (-0.7,1.35) -- (0.3,1.35);
	\node[] at (-0.4,1.55) {\small Index1};	
	
	\draw[->, thick, color=black] (1.2,0.2) -- (1.8,0.2);
	\node[] at (1.4,0.4) {\small last};	
    \draw[->, thick, color=black] (1.2,-0.6) -- (1.8,-0.6);
	\node[] at (1.4,-0.4) {\small first};

\end{tikzpicture}
	\caption{Removing an entry from flowid.}\label{fig:hybridadd}
	\vspace{-0.3cm}
\end{figure}
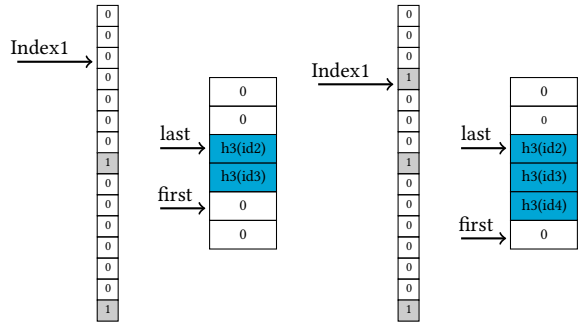
\begin{figure}[ht!]
	\begin{tikzpicture}

	\matrix[tablelong] (mat11) at (2.3,0)
	{
		{\scriptsize 0} \\ {\tiny 0}  \\ |[fill=tudelft]| {\scriptsize h3(id2)} \\ 
		|[fill=tudelft]| {\scriptsize h3(id3)} \\ |[fill=tudelft]| {\scriptsize h3(id4)} \\ {\scriptsize 0} \\
	};
	\matrix[tablelong] (mat11) at (-1.7,0)
	{
		{\scriptsize 0} \\ {\scriptsize 0} \\ |[fill=tudelft]| {\scriptsize h3(id2)} \\ 
		|[fill=tudelft]| {\scriptsize h3(id3)} \\ \scriptsize 0 \\ {\scriptsize 0} \\
	};
	
	\matrix[tablesmall] (mat11) at (-3.5,0)
	{
		{\tiny 0} \\ {\tiny 0} \\ {\tiny 0} \\ {\tiny 0}
		\\ {\tiny 0} \\  {\tiny 0} \\ {\tiny 0}
		\\ |[fill=gray!40]| {\tiny 1} \\  {\tiny 0} \\ {\tiny 0}
		\\ {\tiny 0} \\  {\tiny 0} \\ 	{\tiny 0}	
		\\  {\tiny 0} \\ |[fill=gray!40]| {\tiny 1}  \\
	};

	\matrix[tablesmall] (mat11) at (0.5,0)
	{
		{\tiny 0} \\ {\tiny 0}  \\ {\tiny 0} \\ |[fill=gray!40]| {\tiny 1} \\  {\tiny 0} \\  {\tiny 0} \\ {\tiny 0}
		\\ |[fill=gray!40]| {\tiny 1} \\  {\tiny 0} \\ {\tiny 0}
		\\ {\tiny 0} \\  {\tiny 0} \\ 	{\tiny 0}	
		\\  {\tiny 0} \\ |[fill=gray!40]| {\tiny 1}  \\
	};	
	
	\draw[->, thick, color=black] (-4.7,1.35) -- (-3.7,1.35);
	\node[] at (-4.4,1.55) {\small Index1};	
	\draw[->, thick, color=black] (-2.8,0.2) -- (-2.2,0.2);
	\node[] at (-2.6,0.4) {\small last};	
    \draw[->, thick, color=black] (-2.8,-0.6) -- (-2.2,-0.6);
	\node[] at (-2.6,-0.4) {\small first};

	\draw[->, thick, color=black] (-0.7,1.05) -- (0.3,1.05);
	\node[] at (-0.4,1.25) {\small Index1};	
	
	\draw[->, thick, color=black] (1.2,0.2) -- (1.8,0.2);
	\node[] at (1.4,0.4) {\small last};	
    \draw[->, thick, color=black] (1.2,-1) -- (1.8,-1);
	\node[] at (1.4,-0.8) {\small first};

\end{tikzpicture}
	\caption{Adding an entry to flowid.}\label{fig:hybridadd2}
	\vspace{-0.3cm}
\end{figure}

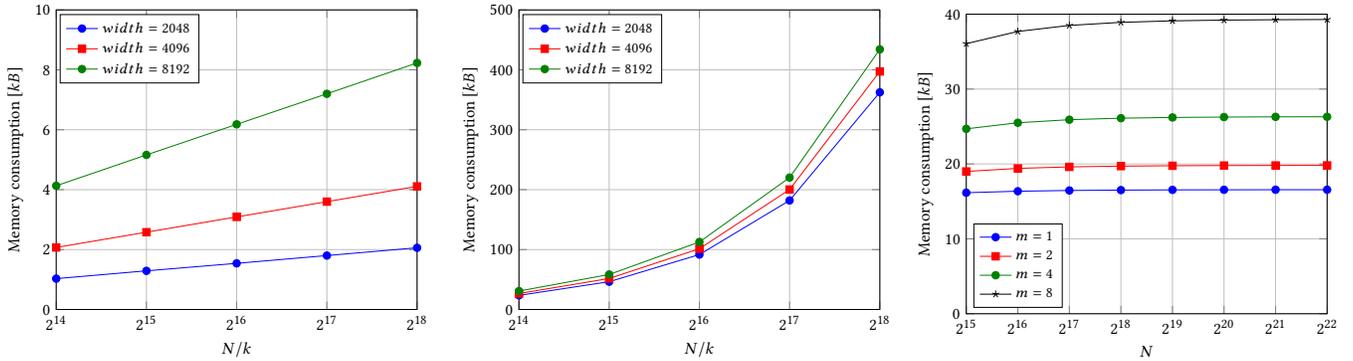
\begin{figure*}[ht!]
	\begin{subfigure}[t]{0.32\textwidth}
	\begin{tikzpicture}[scale = 0.7]
		\begin{semilogxaxis}[
		    ylabel={Memory consumption $\lbrack kB \rbrack$},
		    xlabel={$N/k$},
		    xmin=16384, xmax=262144, ymin=0, ymax=10, grid, log basis x=2,xtick={16384, 32768, 65536, 131072, 262144, 524288},
		    legend style={at={(0.01,0.755)},anchor=south west} 
		    ]
            \addplot+ [blue,mark options={fill=blue}]
    		table [x index=0, y index=1, col sep=comma] 
            {
16384, 1.03
32768, 1.29
65536, 1.54
131072, 1.80
262144, 2.06
524288, 2.31
};
\addplot+ [red,mark options={fill=red}]
    		table [x index=0, y index=1, col sep=comma] 
            {
16384, 2.07
32768, 2.58
65536, 3.09
131072, 3.60
262144, 4.11
524288, 4.63
		    };
            \addplot+ [darkgreen,mark options={fill=darkgreen}]
    		table [x index=0, y index=1, col sep=comma] 
            {
16384, 4.13
32768, 5.16
65536, 6.18
131072, 7.20
262144, 8.23
};

		   \legend{\small $width=2048$, \small $width=4096$, \small $width=8192$}
        \end{semilogxaxis}
	\end{tikzpicture}
	\caption{Memory usage of the initial counting structure.}
	\label{fig:hymem1}
\end{subfigure}\hfill
\begin{subfigure}[t]{0.32\textwidth}
	\begin{tikzpicture}[scale = 0.7]
		\begin{semilogxaxis}[
		    ylabel={Memory consumption $\lbrack kB \rbrack$},
		    xlabel={$N/k$},
		    xmin=16384, xmax=262144, ymin=0, ymax=500, grid, log basis x=2,xtick={16384, 32768, 65536, 131072, 262144},
		    legend style={at={(0.01,0.755)},anchor=south west} 
		    ]
            \addplot+ [blue,mark options={fill=blue}]
    		table [x index=0, y index=1, col sep=comma] 
            {
16384, 23.56
32768, 46.34
65536, 91.66
131072,  182.02
262144, 362.50
};
\addplot+ [red,mark options={fill=red}]
    		table [x index=0, y index=1, col sep=comma] 
            {
16384, 26.64
32768, 51.73
65536, 101.39
131072, 200.21
262144, 397.33
		    };
            \addplot+ [darkgreen,mark options={fill=darkgreen}]
    		table [x index=0, y index=1, col sep=comma] 
            {
16384, 30.76
32768, 58.40
65536, 112.68
131072,  220.20
262144, 434.21
};

		   \legend{\small $width=2048$, \small $width=4096$, \small $width=8192$}
        \end{semilogxaxis}
	\end{tikzpicture}
	\caption{Memory usage of the initial counting structure with the ring of size $N/k$.}
	\label{fig:hymem2}
\end{subfigure}\hfill
\begin{subfigure}[t]{0.32\textwidth}
	\begin{tikzpicture}[scale = 0.7]
		\begin{semilogxaxis}[
		    ylabel={Memory consumption $\lbrack kB \rbrack$},
		    xlabel={$N$},
		    xmin=32768, xmax=4194304, ymin=0, ymax=40, grid, log basis x=2,xtick={16384, 32768, 65536, 131072, 262144, 524288, 1048576, 2097152, 4194304},
		    legend style={at={(0.02,0.02)},anchor=south west} 
		    ]
            \addplot+ [blue,mark options={fill=blue}]
    		table [x index=0, y index=1, col sep=comma] 
            {
16384, 15.75
32768, 16.16
65536, 16.36
131072, 16.46
262144, 16.51
524288, 16.54
1048576, 16.55
2097152, 16.56
4194304, 16.56
};
\addplot+ [red,mark options={fill=red}]
    		table [x index=0, y index=1, col sep=comma] 
            {
16384, 18.19
32768, 19.00
65536, 19.41
131072, 19.61
262144, 19.71
524288, 19.76
1048576, 19.79
2097152, 19.80
4194304, 19.81
		    };
            \addplot+ [darkgreen,mark options={fill=darkgreen}]
    		table [x index=0, y index=1, col sep=comma] 
            {
16384, 23.06
32768, 24.69
65536, 25.50
131072, 25.91
262144, 26.11
524288, 26.21
1048576, 26.26
2097152, 26.29
4194304, 26.30
};

            \addplot+ [black,mark options={fill=black}]
    		table [x index=0, y index=1, col sep=comma] 
            {
16384, 32.81
32768, 36.06
65536, 37.69
131072, 38.50
262144, 38.91
524288, 39.11
1048576, 39.21
2097152, 39.26
4194304, 39.29
};
		   \legend{\small $m=1$, \small $m=2$, \small $m=4$, \small $m=8$}
        \end{semilogxaxis}
	\end{tikzpicture}
	\caption{Memory usage of the flow id array for $th=0.1\%$ and $width=4096$.}
	\label{fig:hymem3}
\end{subfigure}
	\vspace{-0.2cm}
	\caption{Memory consumption of the hybrid window per data-structure.}\label{fig:hybridmem}
	\vspace{-0.4cm}
\end{figure*}
A problem with this window is that smaller flows slowly accumulate in the counting sketch, before finally being removed after they hit $th \cdot N/m$. This increases the number of false positives. To alleviate this problem, we add a smaller pure ring of size $N/m$ to remove packets added to  the initial counting sketch (not from the $th\cdot N/m$ counter) after $N/m$ packets. A heavy hitter typically already hits $th\cdot N / m$ during this time, so the accuracy of its count estimate is not affected by much. However, smaller flows are effectively filtered out.

\textbf{Removing packets from the data-structure.} Every time a new packet arrives, we read the value from the bit array pointing to the oldest received packet ($N$ packets before the packet that is currently processed) using an index (Index1) variable. Index1 increases by one for every processed packet and always points to the oldest entry in the bit array. If the value in the bit array was 1, the count corresponding to that flow is reduced in the counting hash table (data-structure on the right in Fig. \ref{fig:hybrid2}), its bit in the bit array is set to 0, and its flow identifier is removed from the flowid array. Additionally, if an additional pure ring is used a flow needs to be removed from the first counting sketch as described in Sec. \ref{SubSec_ringWindow}.

\textbf{Adding new packets to the data-structure.} 
When a new packet arrives, it is added to the initial counting structure (left data-structure in Fig. \ref{fig:hybrid2}). Every time a flow count of the incoming packet reaches a fraction of the threshold ($th\cdot N/m$) we set the bit in the bit array to 1 and save the flow identifier in the flowid array table. Consequently, we increase the value in the counting hash table (data-structure on the right in Fig. \ref{fig:hybrid2}) by 1. 
To approximate the frequency of an item in the window, we check if this flow ever reached $N\cdot th/m$. If it did, we read the number from the third table and multiply the result with $th/m$. Alternatively, we conclude that the flow is not a heavy hitter and approximate the frequency with the count present in the first counting table. 

\textbf{Memory consumption.}
The total memory consumption of the presented structure is: 
\begin{equation}
\begin{split}
    M_{hybrid} = width_1 \cdot log_2(N/m) + N + width_2\cdot log_2(width3) \\ + width_3 \cdot log_2(m/th)
\end{split}
\end{equation}

The maximum width ($width_2$) of the flowid array can be calculated using the threshold $th$ to detect heavy hitters. When calculating this, we need to consider two consecutive windows of size $N$ (the last $N$ packets that need to be removed and the new $N$ packets that need to be added). In the worst case, if all the counts in the initial sketch have values of $N\cdot th/m -1$ at the same time, and in the next $width1$ packets reach $N\cdot th/m$, they create $width1$ packets that are added to the flowid array. The $2N-width1$ packets left in this window can cause at most $(2N-width1)\cdot m/(N\cdot th)$ packets to reach $N\cdot th/m$.

The memory consumption of the separate data-structures used by the Hybrid window is shown in Fig. \ref{fig:hybridmem}. By adding a smaller pure ring structure, the total memory consumption of the first structure is increased by a factor of 100. Thus, this ring is the main contributor to the overall memory consumption of the first structure (Fig. \ref{fig:hymem1} and Fig. \ref{fig:hymem2}). However, by increasing $m$ to maintain the ratio $N/m$ constant (e.g. $2^{15}$), the total memory consumption of this pure ring, used with a counting sketch with width of $8192$, will be less than $54\;kB$. This number corresponds to just $\approx 3.8\%$ of the memory available per stage on typical programmable hardware ($1.4\;MB$). 

The structure used to count the number of $th\cdot N/m$ occurrences consists of a single table and its memory consumption is similar to the memory consumption of the initial counting structure (Fig. \ref{fig:hymem1}). The total memory consumption of this data-structure is $\leq 20\;kB$ for all analyzed values of $m$, $width$ and $th$ ($m<500$, $width<8192$ and $0.1\% \leq th \leq 1\%$) . 

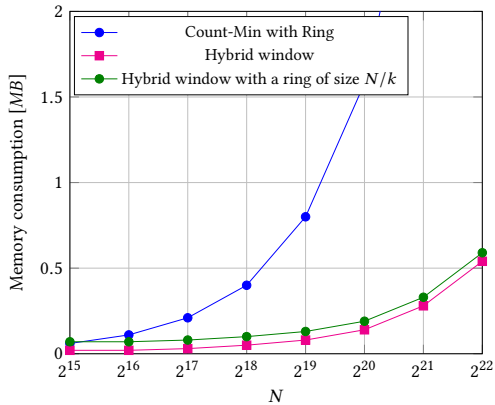
\begin{figure}[ht!]
		\begin{tikzpicture}[scale = 0.8]
		\begin{semilogxaxis}[
		    ylabel={Memory consumption $\lbrack MB \rbrack$},
		    xlabel={$N$},
		    xmin=32767, xmax=4194304, ymin=0, ymax=2, grid, log basis x=2,
		    legend style={at={(0.01,0.755)},anchor=south west}, xtick={32768, 65536, 131072, 262144, 524288, 1048576, 2097152, 4194304},
		    ]
            \addplot+ [blue,mark options={fill=blue}]
    		table [x index=0, y index=1, col sep=comma] 
            {
32768, 0.06
65536, 0.11
131072, 0.21
262144, 0.40
524288, 0.80
1048576, 1.58
2097152, 3.16
4194304, 6.30
		    };
            \addplot+ [magenta,mark options={fill=magenta}]
    		table [x index=0, y index=1, col sep=comma] 
            {		    
32768, 0.02
65536, 0.02
131072, 0.03
262144, 0.05
524288, 0.08
1048576, 0.14
2097152, 0.28
4194304, 0.54
            };
            \addplot+ [darkgreen,mark options={fill=darkgreen}]
    		table [x index=0, y index=1, col sep=comma] 
            {		    
32768, 0.07
65536, 0.07
131072, 0.08
262144, 0.10
524288, 0.13
1048576, 0.19
2097152, 0.33
4194304, 0.59
            };

		   \legend{\small Count-Min with Ring, \small Hybrid window, \small Hybrid window with a ring of size $N/k$ }
        \end{semilogxaxis}
	\end{tikzpicture}
	\vspace{-0.3cm}
	\caption{Comparison of the total memory consumption of the hybrid window ($width1=4096$, $width3=4096$, $th=0.1\%$, $m=N/2^{15}$) and the ring window that is used with a single counting hash table of size $4096$.}\label{fig:hybridmemtotal}
	\vspace{-0.2cm}
\end{figure}

A comparison of the total memory consumption of the Hybrid ring and the Ring window (Sec. \ref{SubSec_ringWindow}) is shown in Fig. \ref{fig:hybridmemtotal}.

 The biggest contributor to the overall memory consumption is the bit array. Its memory consumption scales with $N$ and for the value of $N >2^{23}$ it reaches the hardware limit ($1.4\;MB$).


\section{Evaluation}\label{Sec_Evaluation}

\subsection{Experiment setup}

We implemented and evaluated our approaches on both a Netronome smartNIC as well as by simulation in Python. The 5-tuple consisting of the source IP, destination IP, layer 4 protocol, source port, and the destination port were used as unique flow identifiers. Our python implementation used the same hash function as the one used by Netronome cards (CRC\_CCIT). Different hash functions were created by appending seed values to the flow identifiers. 

\textbf{Traces.} We classified heavy hitters as flows whose frequency was above a threshold $th$ that varied between 0.1\% - 1\%. Packets were obtained from  
10 different traces from an ISP backbone link collected at the Equinix data-center in Chicago in January 2016, made available by CAIDA \cite{caida2016trace}. Each trace is one minute long and contains on average 31 million packets. 

\textbf{Metrics.} We evaluated all our presented counting and sliding window solutions on: (1) percentage of false negatives (percentage of packets that were not reported as belonging to a heavy hitter flow but should have been), and (2) false positives (percentage of packets that were reported as belonging to a heavy hitter flow, but should not have been).

\textbf{Comparison baselines.} We compared our Gated sketch against the Count-Min sketch with the same memory consumption. The Count-Min sketch was chosen as a baseline algorithm. We compared our sliding window approaches to simply periodically resetting all registers and setting them to 0, since we are not aware of any other P4 solution that implements sliding windows.

\subsection{Counting sketches: Accuracy}

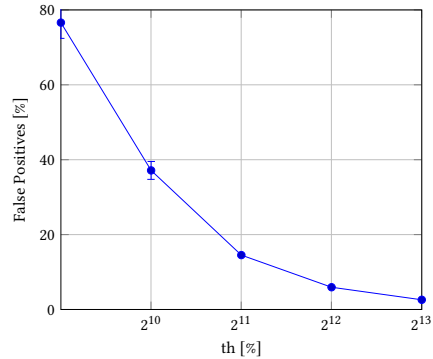
\begin{figure}[h!]
\centering
	\begin{tikzpicture}[scale = 0.7]
	    \begin{semilogxaxis}[
		    xlabel={th $\lbrack \% \rbrack$},
		    ylabel={False Positives $\lbrack \% \rbrack$},
	        xmin=512, xmax=8192, ymin=0, ymax=80, grid,log basis x=2,
	        legend style={at={(0.97,0.1)},anchor=south east}
	        ]
		    \addplot+ [error bars/.cd, y explicit, y dir=both]
    		table [x index=0, y index=1, col sep=comma,
	    	y error plus index=2,
		    y error minus index=2
		    ] 
            {
512 ,76.608443, 4.204812
1024 ,37.138688, 2.389914
2048 ,14.526260, 0.283720
4096 ,5.940959, 0.129456
8192 ,2.588794, 0.056311
            };
	    \end{semilogxaxis}
	\end{tikzpicture}
\caption{Accuracy of the Count-Min sketch with the Ring window for different widths by classifying using 10 different CAIDA traces. Window size $2^{16}$, depth 3, threshold 0.1\% and confidence interval 95\%.}\label{fig:mininet}
\end{figure}

\textbf{Count-Min sketch.} The number of false positives mostly depends on the width of the sketch (Fig. \ref{fig:mininet}). Increasing the width of the Count-Min sketch reduces the number of hash collisions and, as a direct result, reduces the count overestimation. False positives also decrease with the depth of the sketch, but not significantly. 

\begin{figure*}[tbh!]
\centering
\begin{subfigure}[t]{0.32\textwidth}
	\begin{tikzpicture}[scale = 0.7]
		\begin{axis}[
		    xlabel={th0 $\lbrack \# packets \rbrack$},
		    ylabel={False Positives $\lbrack \% \rbrack$},
		    xmin=10, xmax=60, ymin=0, ymax=56, grid,
		    legend style={at={(0.6,0.8)},anchor=south west} 
		    ]
		    \addplot+ [error bars/.cd, y explicit, y dir=both]
    		table [x index=0, y index=1, col sep=comma,
	    	y error plus index=2,
		    y error minus index=2
		    ] 
		    {10,55.022908, 0.598707
            15,42.061452, 0.549830
            20,32.738628, 0.472933
            25,26.292455, 0.365483
            30,21.640898, 0.303668
            35,18.097152, 0.255852
            40,15.266339, 0.223422
            45,12.901579, 0.198021
            50,10.855398, 0.177750
            55, 8.989451, 0.161652
            60, 7.252205, 0.145267
            };
		    \addplot+ [error bars/.cd, y explicit, y dir=both,]
    		table [x index=0, y index=1, col sep=comma,
	    	y error plus index=2,
		    y error minus index=2
		    ] 
		    {
		    10,14.226707,0.246305
            15,14.226707,0.246305
            20,14.226707,0.246305
            25,14.226707,0.246305
            30,14.226707,0.246305
            35,14.226707,0.246305
            40,14.226707,0.246305
            45,14.226707,0.246305
            50,14.226707,0.246305
            55,14.226707,0.246305
            60,14.226707,0.246305
            };
	        \legend{Gated, Count-Min}
		\end{axis}
	\end{tikzpicture}
	\caption{Comparison of the percentage of false positives for a Gated sketch (with width\_0 = 4096, width\_1=2048) and the Count-Min sketch with the width 2048 and depth 3.}
	\label{fig:gatedev1}
\end{subfigure}\hfill
\begin{subfigure}[t]{0.32\textwidth}
	\begin{tikzpicture}[scale = 0.7]
		\begin{axis}[
		    xlabel={th0 $\lbrack \# packets \rbrack$},
		    ylabel={Number of packets $\lbrack \% \rbrack$},
		    xmin=5, xmax=64, ymin=0, ymax=100, grid,
		    legend style={at={(0.03,0.05)},anchor=south west} 
		    ]
		    \addplot+ [error bars/.cd, y explicit, y dir=both,]
    		table [x index=0, y index=1, col sep=comma,
	    	y error plus index=2,
		    y error minus index=2
		    ] 
		    {
			5,100,0
			10,100,0
			15,100,0
			20,100,0
			25,100,0
			30,100,0
			35,100,0
			40,100,0
			45,100,0
			50,100,0
			55,100,0
			60,100,0
			64,100,0
            };		    
		    \addplot+ [error bars/.cd, y explicit, y dir=both]
    		table [x index=0, y index=1, col sep=comma,
	    	y error plus index=2,
		    y error minus index=2
		    ] 
		    {5,92.443996, 0.045620  
            10,77.205053, 0.070844 
            15,63.739875, 0.090076 
            20,54.057237, 0.139405 
            25,47.127682, 0.213163 
            30,41.932126, 0.283222 
            35,37.838321, 0.352427 
            40,34.485649, 0.414715 
            45,31.658521, 0.465374 
            50,29.224796, 0.502192 
            55,27.064557, 0.522936 
            60,25.137846, 0.532148 
            64,23.765458, 0.539637 
            };
	        \legend{Count-Min, Gated}
		\end{axis}
	\end{tikzpicture}
	\caption{Comparison of the percentage of packets processed by the second stage of the Gated sketch (with width\_0 = 4096, width\_1=2048) and the Count-Min sketch with the width 2048 and depth 3.}
	\label{fig:gatedev2}
\end{subfigure}\hfill
\begin{subfigure}[t]{0.32\textwidth}
	\begin{tikzpicture}[scale = 0.7]
		\begin{semilogxaxis}[
		    xlabel={$width_1/width\_0$},
		    ylabel={False positives $\lbrack \% \rbrack$},
		    xmin=0.0625, xmax=1, ymin=2.5, ymax=15.5, grid,log basis x=2,
		    legend style={at={(0.41,0.5)},anchor=south west} 
		    ]
		    \addplot+ [error bars/.cd, y explicit, y dir=both]
    		table [x index=0, y index=1, col sep=comma,
	    	y error plus index=2,
		    y error minus index=2
		    ] 
            {
            0.0625,2.808, 0.063
            0.125,2.695, 0.059
            0.25,2.634, 0.058
            0.5,2.607, 0.057
            1,2.589, 0.056
            };
		    \addplot+ [error bars/.cd, y explicit, y dir=both]
    		table [x index=0, y index=1, col sep=comma,
	    	y error plus index=2,
		    y error minus index=2
		    ] 
            {
            0.0625 ,6.505, 0.145
            0.125 ,6.249, 0.138
            0.25,6.077, 0.133
            0.5,5.984, 0.131
            1,5.941, 0.129
            };  
		    \addplot+ [error bars/.cd, y explicit, y dir=both]
    		table [x index=0, y index=1, col sep=comma,
	    	y error plus index=2,
		    y error minus index=2
		    ] 
            {
            0.125,15.272, 0.269
            0.25,14.774, 0.258
            0.5,14.422, 0.253
            1,14.227, 0.246
            };    
	        \legend{Gated($width0=8192$), Gated($width0=4096$), Gated($width0=2048$)}
	        
		\end{semilogxaxis}
	\end{tikzpicture}
	\caption{Percentage of false positives of a Gated sketch for different values of $width\_1$ and $width_0$.}
	\label{fig:gatedev3}
\end{subfigure}\hfill
\caption{Comparison of the Gated sketch and the Count-Min sketch with Ring window for different widths by classifying packets of 10 different CAIDA traces. Window size 65536, threshold 0.1\% (65 packets), and confidence interval 95\%.}\label{fig:gatedev}
\vspace{-0.3cm}
\end{figure*}
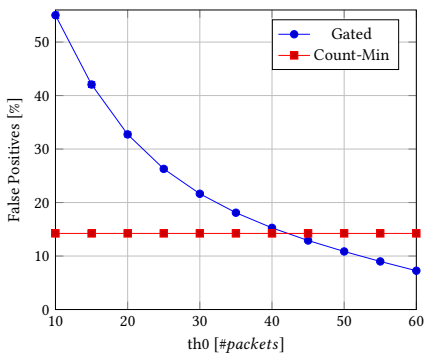
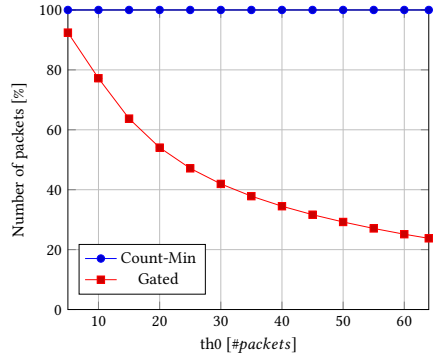
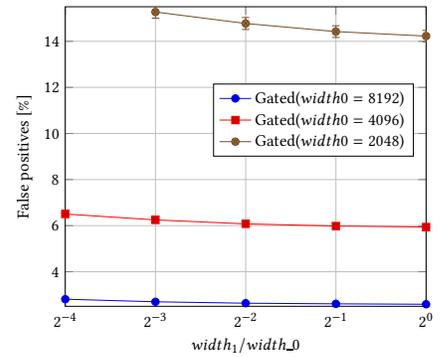

\textbf{Gated sketch.} Gated sketch outperforms the previously implemented Count-Min sketch in both accuracy and memory usage. Its accuracy mostly depends on the thresholds and widths used at each stage. This is especially true for the first threshold ($th0$) and width ($width0$) since they have a significant influence on the number of packets processed in the later stages (as can be seen in Fig. \ref{fig:gatedev2}). 

A higher $th0$ reduces the number of collisions in deeper stages, increasing accuracy and decreasing the number of false positives (Fig. \ref{fig:gatedev1}). Additionally, by increasing $width0$ the number of packets processed in the deeper stages due to collisions is reduced (as the load factor is reduced similarly to the Count-Min sketch). As a consequence, only heavy hitter flows and smaller flows that collide with them in the first stage are processed in the deeper stages.

Since only a small fraction of packets is processed by the deeper stages, their width can be reduced without losing much accuracy. This reduces the overall memory consumption, making it possible to trade-off the width of the deeper stages for the width of the first stage (Fig. \ref{fig:gatedev3}). Additionally, smaller flows that pass through the first stage (due to collisions) are filtered in the deeper stages, since the probability of them colliding with another heavy hitter flow in all deeper stages is reduced.

For example (see Fig. \ref{fig:gatedev1}) for a window of size 65536 and a threshold equal to 0.1\% (65 packets) the percentage of false positives of a Gated sketch with a $width_0$ of 4096 and $width_1$ of 2048 varies between $\approx 55\%$ (threshold $th0$ set to 10) and $\approx 7\%$ (threshold $th0$ set to 60). At the same time, the Count-Min sketch with a depth of 3 and width of 4096 (thus, with the same number of count entries) does not achieve a lower percentage of false positives than $14\%$. 

\subsection{Sliding window: Accuracy}

\textbf{Flushing. } We tested the accuracy of the counting sketches when the structure was flushed every $N$ packets (Fig. \ref{fig:flushingcountmin}), as this is the most commonly used method found in the literature (\cite{hashpipe}) to clear data-structures. This method is used as the baseline to compare our solutions to.
\begin{figure}[h!]
\centering
	\begin{tikzpicture}[scale = 0.85]
		\begin{semilogxaxis}[
		    ylabel={Probability of false negatives $\lbrack \% \rbrack$},
		    xlabel={$N$},
		    xmin=32767, xmax=524288, ymin=0, ymax=10, grid, log basis x=2,	        xtick={32768, 65536, 131072, 262144, 524288},
		    legend style={at={(0.01,0.01)},anchor=south west} 
		    ]
            \addplot+ [error bars/.cd, y explicit, y dir=both]
    		table [ 
    		x index=0, y index=1, col sep=comma, 
    		y error plus index=2,
		    y error minus index=2,
    		] 
            {
32768 ,6.135459, 0.105301
65536 ,5.299941, 0.130195
131072 ,5.702888, 0.148184
262144 ,5.179037, 0.167440
524288 ,4.737645, 0.174980
1048576 ,4.227512, 0.165487
		    };
		    
       \end{semilogxaxis}
	\end{tikzpicture}
\caption{Comparison of the probability of false negatives for the Count-Min sketch with $width=8192$, $depth=3$ and $th=0.1\%$ and which is reset every $N$ packets. Confidence interval is equal to  95\%.}\label{fig:flushingcountmin}
\end{figure}
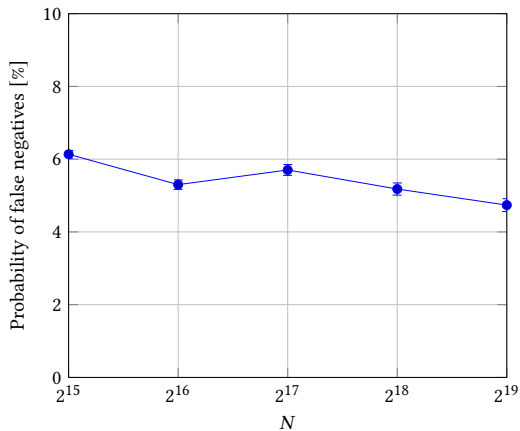
\begin{figure*}[tbh!]
\centering
\begin{subfigure}[t]{0.32\textwidth}
	\begin{tikzpicture}[scale = 0.7]
	    \begin{axis}[
		    xlabel={th $\lbrack \% \rbrack$},
		    ylabel={False Positives $\lbrack \% \rbrack$},
	        xmin=0.1, xmax=1, ymin=0, ymax=60, grid,
	        legend style={at={(0.98,0.1)},anchor=south east}
	        ]
		    \addplot+ [error bars/.cd, y explicit, y dir=both]
    		table [x index=0, y index=1, col sep=comma,
	    	y error plus index=2,
		    y error minus index=2
		    ] 
            {
0.1 ,1.347721, 0.662842
0.2 ,0.831355, 0.047013
0.3 ,0.495844, 0.047773
0.4 ,0.314885, 0.026376
0.5 ,0.236547, 0.039804
0.6 ,0.152437, 0.026754
0.7 ,0.130844, 0.052725
0.8 ,0.089473, 0.025504
0.9 ,0.065290, 0.01109
1,0.056001, 0.009251
            };  	        
		    \addplot+ [error bars/.cd, y explicit, y dir=both]
    		table [x index=0, y index=1, col sep=comma,
	    	y error plus index=2,
		    y error minus index=2
		    ] 
            {
0.1 ,45.950535, 3.105452
0.2 ,50.481409, 2.355999
0.3 ,50.779891, 2.426164
0.4 ,49.841561, 2.405826
0.5 ,48.882584, 1.868719
0.6 ,47.644199, 1.689558
0.7 ,46.544406, 1.413550
0.8 ,45.512199, 1.290069
0.9 ,44.402803, 1.178060
1 ,43.322991, 1.078084
            };            
	        
	        \legend{\small Ring(depth=3), \small Sequential(depth = 3)}
	    \end{axis}
	\end{tikzpicture}
	\caption{Percentage of false positives.}
	\label{fig:falseposcountminseq1}
\end{subfigure}\hfill
\begin{subfigure}[t]{0.32\textwidth}
	\begin{tikzpicture}[scale = 0.7]
	    \begin{semilogxaxis}[
		    xlabel={N $\lbrack \#packeta \rbrack$},
		    ylabel={False Positives $\lbrack \% \rbrack$},
	        xmin=32768, xmax=524288, ymin=0, ymax=60, grid,log basis x=2,	        xtick={32768, 65536, 131072, 262144, 524288},
	        legend style={at={(0.98,0.82)},anchor=south east}
	        ]
		    \addplot+ [error bars/.cd, y explicit, y dir=both]
    		table [x index=0, y index=1, col sep=comma,
	    	y error plus index=2,
		    y error minus index=2
		    ] 
            {
32768 ,2.860026, 0.159565
65536 ,2.656059, 0.286886
131072 ,2.343245, 0.098228
262144 ,2.202648, 0.135495
524288 ,2.120076, 0.206652
            };

		    \addplot+ [error bars/.cd, y explicit, y dir=both]
    		table [x index=0, y index=1, col sep=comma,
	    	y error plus index=2,
		    y error minus index=2
		    ] 
            {

32768 ,49.339096, 0.744473
65536, 48.470727, 2.470357
131072 ,45.868131, 3.447911
262144 ,41.819786, 4.461201
524288 ,36.370660, 5.610863
1048576 ,47.196031, 3.252356
};
	        
	        \legend{\small Ring(depth=3), \small Sequential(depth = 3)}
	    \end{semilogxaxis}
	\end{tikzpicture}
	\caption{Percentage of false positives.}
	\label{fig:falseposcountminseq2}
\end{subfigure}\hfill
\begin{subfigure}[t]{0.32\textwidth}
	\begin{tikzpicture}[scale = 0.7]
	    \begin{semilogxaxis}[
		    xlabel={sketch width $\lbrack bits \rbrack$},
		    ylabel={False Positives $\lbrack \% \rbrack$},
	        xmin=512, xmax=8192, ymin=0, ymax=80, grid,log basis x=2,
	        xtick={512, 1024, 2048, 4096, 8192},
	        legend style={at={(0.98,0.82)},anchor=south east}
	        ]
		    \addplot+ [error bars/.cd, y explicit, y dir=both]
    		table [x index=0, y index=1, col sep=comma,
	    	y error plus index=2,
		    y error minus index=2
		    ] 
            {
512 ,75.707020, 3.223901
1024 ,36.016680, 2.557665
2048 ,14.400259, 0.852360
4096 ,6.146664, 0.436606
8192 ,2.656059, 0.286886
            };

		    \addplot+ [error bars/.cd, y explicit, y dir=both]
    		table [x index=0, y index=1, col sep=comma,
	    	y error plus index=2,
		    y error minus index=2
		    ] 
            {            
512 ,52.797529, 4.342634
1024 ,50.733539, 4.507509
2048 ,48.931486, 2.814938
4096 ,48.231551, 2.770735
8192 ,48.458880, 2.471006
            };
	        
	        \legend{\small Ring(depth=3), \small Sequential(depth = 3)}
	    \end{semilogxaxis}
	\end{tikzpicture}
	\caption{Percentage of false positives.}
	\label{fig:falseposcountminseq3}
\end{subfigure}\hfill
\begin{subfigure}[t]{0.32\textwidth}
	\begin{tikzpicture}[scale = 0.7]
	    \begin{axis}[
		    xlabel={th $\lbrack \% \rbrack$},
		    ylabel={False Negatives $\lbrack \% \rbrack$},
	        xmin=0.1, xmax=1, ymin=0, ymax=0.01, grid,
	        legend style={at={(0.98,0.82)},anchor=south east}
	        ]
		    \addplot+ [error bars/.cd, y explicit, y dir=both]
    		table [x index=0, y index=1, col sep=comma,
	    	y error plus index=2,
		    y error minus index=2
		    ] 
            {
0.1 ,0, 0
0.2 ,0, 0
0.3 ,0, 0
0.4 ,0, 0
0.5 ,0, 0
0.6 ,0, 0
0.7 ,0, 0
0.8 ,0, 0
0.9 ,0, 0
1 ,0, 0
};
		    \addplot+ [error bars/.cd, y explicit, y dir=both]
    		table [x index=0, y index=1, col sep=comma,
	    	y error plus index=2,
		    y error minus index=2
		    ] 
            {
0.1 ,0.005163, 0.005374
0.2 ,0.002824, 0.000233
0.3 ,0.001446, 0.000548
0.4 ,0.000961, 0.000407
0.5 ,0.000773, 0.000149
0.6 ,0.000725, 0.000191
0.7 ,0.000635, 0.000079
0.8 ,0.000576, 0.000110
0.9 ,0.000494, 0.000260
1 ,0.000491, 0.000194
};

	        \legend{\small Ring(depth=3), \small Sequential(depth = 3)}
	    \end{axis}
	\end{tikzpicture}
	\caption{Percentage of false negatives.}
	\label{fig:falsenegativecountmin1}
\end{subfigure}
\begin{subfigure}[t]{0.32\textwidth}
	\begin{tikzpicture}[scale = 0.7]
	    \begin{semilogxaxis}[
		    xlabel={sketch width $\lbrack w \rbrack$},
		    ylabel={False Negatives $\lbrack \% \rbrack$},
	        xmin=32768, xmax=524288, ymin=0, ymax=0.03, grid,log basis x=2,
	        xtick={32768, 65536, 131072, 262144, 524288},
	        legend style={at={(0.98,0.82)},anchor=south east}
	        ]
		    \addplot+ [error bars/.cd, y explicit, y dir=both]
    		table [x index=0, y index=1, col sep=comma,
	    	y error plus index=2,
		    y error minus index=2
		    ] {
		    32768, 0, 0
65536, 0, 0
131072, 0, 0
262144, 0, 0
524288, 0, 0
1048576, 0, 0
};
		    \addplot+ [error bars/.cd, y explicit, y dir=both]
    		table [x index=0, y index=1, col sep=comma,
	    	y error plus index=2,
		    y error minus index=2
		    ] 
            {
32768 ,0.009620, 0.002073
65536, 0.011149, 0.000689
131072 ,0.013941, 0.006389
262144 ,0.015583, 0.000527
524288 ,0.020720, 0.005794
1048576 ,0.012071, 0.000084
	        };

	        \legend{\small Ring(depth=3), \small Sequential(depth = 3)}
	    \end{semilogxaxis}
	\end{tikzpicture}
	\caption{Percentage of false negatives.}
	\label{fig:falsenegativecountmin2}
\end{subfigure}
\begin{subfigure}[t]{0.32\textwidth}
	\begin{tikzpicture}[scale = 0.7]
	    \begin{semilogxaxis}[
		    xlabel={sketch width $\lbrack bits \rbrack$},
		    ylabel={False Negatives $\lbrack \% \rbrack$},
	        xmin=512, xmax=8192, ymin=0, ymax=0.3, grid,log basis x=2,
	        xtick={512, 1024, 2048, 4096, 8192},
	        legend style={at={(0.98,0.82)},anchor=south east}
	        ]
		    \addplot+ [error bars/.cd, y explicit, y dir=both]
    		table [x index=0, y index=1, col sep=comma,
	    	y error plus index=2,
		    y error minus index=2
		    ] 
            {            
            
512 ,0, 0
1024 ,0, 0
2048 ,0, 0
4096 ,0, 0
8192 ,0, 0
            };
		    \addplot+ [error bars/.cd, y explicit, y dir=both]
    		table [x index=0, y index=1, col sep=comma,
	    	y error plus index=2,
		    y error minus index=2
		    ] 
            {            
            
512 ,0.239574, 0.037813
1024 ,0.102617, 0.021169
2048 ,0.041333, 0.005951
4096 ,0.019164, 0.001154
8192 ,0.011149, 0.000689
            };
	        
	        \legend{\small Ring(depth=3), \small Sequential(depth = 3)}
	    \end{semilogxaxis}
	\end{tikzpicture}
	\caption{Percentage of false negatives.}
	\label{fig:falsenegativecountmin3}
\end{subfigure}
\caption{Accuracy of the Count-Min sketch with the sequential and ring window for different values of the widths by classifying UDP and TCP packets of 10 different CAIDA traces. Window size 65536, threshold 0.1\% (65 packets) and confidence interval 95\%.}\label{fig:countminevaluation}
\vspace{-0.5cm}
\end{figure*}
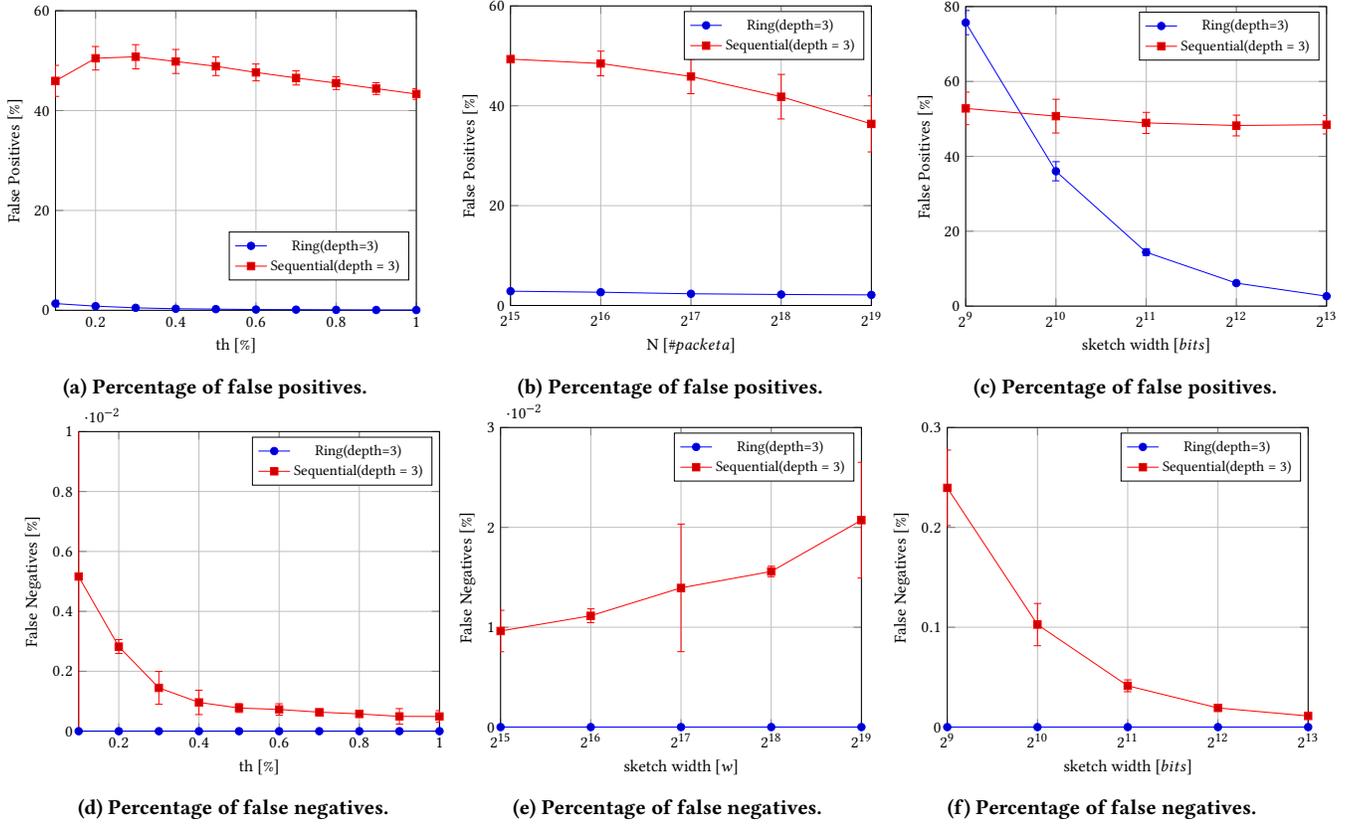
If used with a Count-Min sketch with a width of $8192$, the probability of false positives is low (less than 1.2\%) and decreases with the increase of the window size $N$ (to 0.6\% for $N=2^{19}$). The reason for this is that the used threshold to identify the heavy hitters, configured as $N\cdot 0.1\%$, increases with $N$ (e.g., 524 packets for $N=2^{19}$ compared to 32 packets for  $N=2^{15}$). However, the probability of false negatives can be significant and depends on the window size $N$. For lower values of $N$, the counting structure is reset more frequently and identified heavy hitters forgotten more often thereby increasing the probability of false negatives from 4.2\% for $N=2^{19}$ to 6.2\% for $N=2^{15}$. 

\textbf{Ring window.} The ring window has the best accuracy among all the analyzed solutions. The probability of false negatives is equal to 0 for all the analyzed values of width, $N$, and $th$. This is expected as the counting sketch (e.g. Count-Min sketch) only stores the values of the last $N$ packets. Thus, counts can only be overestimated, and never underestimated. 


The probability of false positives is mostly influenced by the width of the sketch (Fig. \ref{fig:falseposcountminseq3}). A larger width decreases the number of hash collisions, and with it the number of false positives. Similarly, increased depth reduces the number of false positives. However, the influence of depth is less significant than that of the width.


\textbf{Sequential window.} This solution performs worst of all analyzed solutions. 
\begin{figure*}[tbh!]
\centering
\input{figures/flushingev1.tex}
\caption{Comparison of the Sequential Flushing approach for different depths and widths using a Count-Min sketch by classifying UDP and TCP packets using 10 different CAIDA traces. Confidence interval is 95\%.}\label{fig:flushingseq}
\vspace{-0.4cm}
\end{figure*}
In contrast to the ring implementation, many entries in the tables can be 0 (depending on the width of the table) causing the total number of counts increased per window to be larger than the total amount of counts reduced. As these counts are never removed, accuracy decreases over time resulting in a significant number of false positives (between 40\% and 55\%) even for large width values. 
Moreover, the probability of false negatives is not equal to 0 (as in the Ring window). This is because the entries that are reduced by the sequential window, do not necessarily correspond to the oldest packet of the window. However, the percentage of false negatives is significantly lower than the percentage of false positives ($\leq 2\%$).

\textbf{Sequential flushing.} The probability of false positives and false negatives of the sequential flushing approach predominantly depends on two parameters: (1) the depth and (2) the width of the counting sketch. 

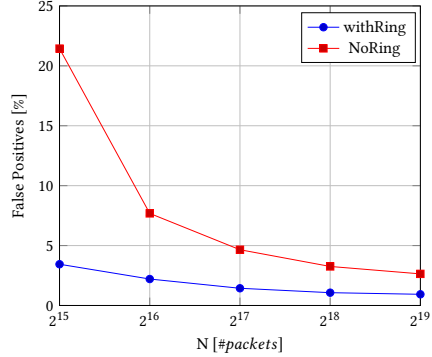
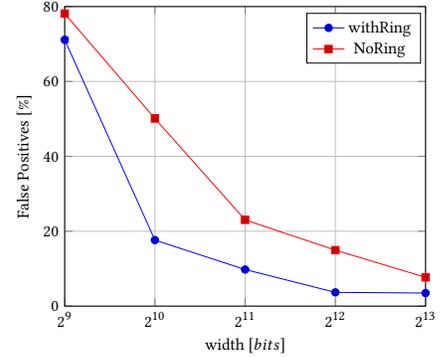
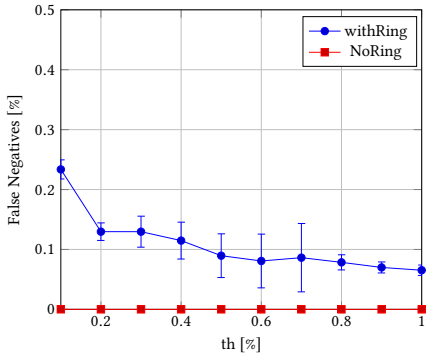
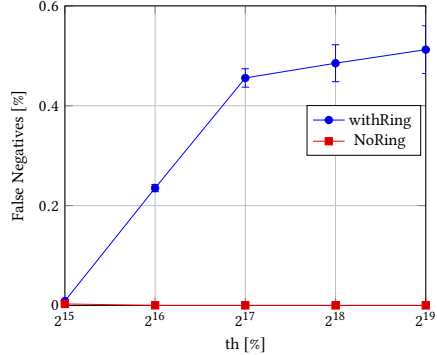
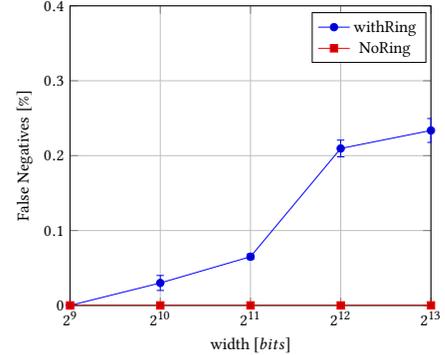
\begin{figure*}[tbh!]
\centering
\begin{subfigure}[t]{0.32\textwidth}
	\begin{tikzpicture}[scale = 0.7]
		\begin{axis}[
		    xlabel={th $\lbrack \% \rbrack$},
		    ylabel={False Positives $\lbrack \% \rbrack$},
		    xmin=0.1, xmax=1, ymin=0, ymax=10, grid,
		    legend style={at={(0.67,0.81)},anchor=south west} 
		    ]
		    
		    \addplot+ [error bars/.cd, y explicit, y dir=both]
    		table [x index=0, y index=1, col sep=comma,
	    	y error plus index=2,
		    y error minus index=2
		    ] 
		    {
            0.1, 2.207989, 0.072685	    
            0.2,0.796775, 0.155894
            0.3, 0.662015, 0.212127
            0.4, 0.486756, 0.081112
            0.5, 0.415734, 0.149585
            0.6, 0.254045, 0.108442
            0.7, 0.172679, 0.076152
            0.8, 0.149831, 0.029290
            0.9, 0.101960, 0.014548
            1 ,0.091627, 0.010492
		    };
		    \addplot+ [error bars/.cd, y explicit, y dir=both]
    		table [x index=0, y index=1, col sep=comma,
	    	y error plus index=2,
		    y error minus index=2
		    ] 
		    {
0.1 ,7.685303, 0.159204
0.2 ,3.909246, 0.305679
0.3 ,3.421828, 0.388851
0.4 ,2.517636, 0.086130
0.5 ,2.049212, 0.411783
0.6 ,1.612480, 0.431422
0.7 ,1.435420, 0.500790
0.8 ,1.309323, 0.549620
0.9 ,1.131009, 0.398756
1 ,1.006095, 0.282858
		    };

	        \legend{withRing, NoRing}
		\end{axis}
	\end{tikzpicture}
	\caption{Percentage of false positives for $width=8192$ and $N=2^{16}$.}
	\label{fig:hybrid1ev}
\end{subfigure}\hfill
\begin{subfigure}[t]{0.32\textwidth}
	\begin{tikzpicture}[scale = 0.7]
		\begin{semilogxaxis}[
		    xlabel={N $\lbrack \#packets \rbrack$},
		    ylabel={False Positives $\lbrack \% \rbrack$},
		    xmin=32768, xmax=524288, ymin=0, ymax=25, grid,log basis x=2,	        xtick={32768, 65536, 131072, 262144, 524288},
		    legend style={at={(0.67,0.81)},anchor=south west} 
		    ]
		    
		    \addplot+ [error bars/.cd, y explicit, y dir=both]
    		table [x index=0, y index=1, col sep=comma,
	    	y error plus index=2,
		    y error minus index=2
		    ] 
		    {
32768, 3.442237, 0.109967
65536 ,2.207989, 0.072685
131072 ,1.435850, 0.063805
262144 ,1.068832, 0.058538
524288 ,0.933639, 0.050063
1048576 ,0.861823, 0.056565
            };
		    \addplot+ [error bars/.cd, y explicit, y dir=both]
    		table [x index=0, y index=1, col sep=comma,
	    	y error plus index=2,
		    y error minus index=2
		    ] 
		    {
32768,21.439766, 0.252720
65536,7.685303, 0.159204
131072,4.653394, 0.161707
262144,3.257055, 0.158190
524288,2.638400, 0.146096
    };		  
	        \legend{withRing, NoRing}
		\end{semilogxaxis}
	\end{tikzpicture}
	\caption{Percentage of false positives for $width=8192$ and $th=0.1\%$.}
	\label{fig:hybrid2ev}
\end{subfigure}\hfill
\begin{subfigure}[t]{0.32\textwidth}
	\begin{tikzpicture}[scale = 0.7]
		\begin{semilogxaxis}[
		    xlabel={width $\lbrack bits \rbrack$},
		    ylabel={False Positives $\lbrack \% \rbrack$},
		    xmin=512, xmax=8192, ymin=0, ymax=80, grid,log basis x=2,xtick={512, 1024, 2048, 4096, 8192},
		    legend style={at={(0.67,0.81)},anchor=south west} 
		    ]
		    
		    \addplot+ [error bars/.cd, y explicit, y dir=both]
    		table [x index=0, y index=1, col sep=comma,
	    	y error plus index=2,
		    y error minus index=2
		    ] 
		    {
512 ,71.127128, 0.793456
1024,17.634193, 0.123451
2048 ,9.772084, 0.203361
4096 ,3.672193, 0.126335
8192, 3.483237, 0.171841
            };
		    \addplot+ [error bars/.cd, y explicit, y dir=both]
    		table [x index=0, y index=1, col sep=comma,
	    	y error plus index=2,
		    y error minus index=2
		    ] 
		    {
512 ,78.123859, 0.694804
1024 ,50.098253, 0.736267
2048 ,23.029076, 0.332190
4098 ,14.943116, 0.307676
8192 ,7.685303, 0.159204
 };          
		  
	        \legend{withRing, NoRing}
		\end{semilogxaxis}
	\end{tikzpicture}
	\caption{Percentage of false positives for $width=8192$ and $th=0.1\%$.}
	\label{fig:hybrid3ev}
\end{subfigure}\hfill
\begin{subfigure}[t]{0.32\textwidth}
	\begin{tikzpicture}[scale = 0.7]
		\begin{axis}[
		    xlabel={th $\lbrack \% \rbrack$},
		    ylabel={False Negatives $\lbrack \% \rbrack$},
		    xmin=0.1, xmax=1, ymin=0, ymax=0.5, grid,
		    legend style={at={(0.67,0.81)},anchor=south west} 
		    ]
		    \addplot+ [error bars/.cd, y explicit, y dir=both]
    		table [x index=0, y index=1, col sep=comma,
	    	y error plus index=2,
		    y error minus index=2
		    ] 
		    {
            0.1,0.233617, 0.015948
            0.2,0.129702, 0.014653
            0.3,0.129661, 0.025885
            0.4,0.114786, 0.030776
            0.5,0.089593, 0.036495
            0.6,0.080766, 0.044861
            0.7,0.086294, 0.057050
            0.8,0.078474, 0.012647
            0.9,0.069971, 0.009179
            1,  0.065319, 0.008652
            };
  		    \addplot+ [error bars/.cd, y explicit, y dir=both]
    		table [x index=0, y index=1, col sep=comma,
	    	y error plus index=2,
		    y error minus index=2
		    ] 
		    {
0.1 ,0.000078, 0.000040
0.2 ,0.000017, 0.000050
0.3 ,0.000000, 0.000000
0.4 ,0.000000, 0.000000
0.5 ,0.000023, 0.000074
0.6 ,0.000000, 0.000000
0.7 ,0.000015, 0.000048
0.8 ,0.000000, 0.000000
0.9 ,0.000000, 0.000000
1 ,0.000000, 0.000000
};            

	        \legend{withRing, NoRing}
		\end{axis}
	\end{tikzpicture}
	\caption{Percentage of false negatives for $width=8192$, $m=2$ and $N=2^{16}$.}
	\label{fig:hybrid4ev}
\end{subfigure}\hfill
\begin{subfigure}[t]{0.32\textwidth}
	\begin{tikzpicture}[scale = 0.7]
		\begin{semilogxaxis}[
		    xlabel={th $\lbrack \% \rbrack$},
		    ylabel={False Negatives $\lbrack \% \rbrack$},
		    xmin=32768, xmax=524288, ymin=0, ymax=0.6, grid,log basis x=2,	        xtick={32768, 65536, 131072, 262144, 524288},
		    legend style={at={(0.67,0.5)},anchor=south west} 
		    ]
		    \addplot+ [error bars/.cd, y explicit, y dir=both]
    		table [x index=0, y index=1, col sep=comma,
	    	y error plus index=2,
		    y error minus index=2
		    ] 
		    {
            32768,0.008891, 0.002744
65536 ,0.235104, 0.007229
131072 ,0.455641, 0.018508
262144 ,0.485168, 0.036990
524288 ,0.512339, 0.047796
1048576 ,0.497270, 0.064955
            };
		    \addplot+ [error bars/.cd, y explicit, y dir=both]
    		table [x index=0, y index=1, col sep=comma,
	    	y error plus index=2,
		    y error minus index=2
		    ] 
		    {
            
32768,0.002978, 0.000334
65536,0.000078, 0.000040
131072,0.000000, 0.000000
262144,0.000000, 0.000000
524288,0.000000, 0.000000
};
          
	        \legend{withRing, NoRing}
		\end{semilogxaxis}
	\end{tikzpicture}
	\caption{Percentage of false negatives for $width=8192$ and $th=0.1\%$.}
	\label{fig:hybrid5ev}
\end{subfigure}\hfill
\begin{subfigure}[t]{0.32\textwidth}
	\begin{tikzpicture}[scale = 0.7]
		\begin{semilogxaxis}[
		    xlabel={width $\lbrack bits \rbrack$},
		    ylabel={False Negatives $\lbrack \% \rbrack$},
		    xmin=512, xmax=8192, ymin=0, ymax=0.4, grid,log basis x=2,xtick={512, 1024, 2048, 4096, 8192},
		    legend style={at={(0.67,0.81)},anchor=south west} 
		    ]
		    \addplot+ [error bars/.cd, y explicit, y dir=both]
    		table [x index=0, y index=1, col sep=comma,
	    	y error plus index=2,
		    y error minus index=2
		    ] 
		    {
512 ,0.000015, 0.000022
1024 ,0.030000, 0.010000
2048 ,0.065064, 0.003667
4096 ,0.209586, 0.011167
8192, 0.233617, 0.015948
            };
		    
		    \addplot+ [error bars/.cd, y explicit, y dir=both]
    		table [x index=0, y index=1, col sep=comma,
	    	y error plus index=2,
		    y error minus index=2
		    ] 
		    {
512 ,0.000000, 0.000000
1024 ,0.000000, 0.000000
2048 ,0.000000, 0.000000
4096 ,0.000000, 0.000000
8192, 0.000078, 0.000040
            };

	        \legend{withRing, NoRing}
		\end{semilogxaxis}
	\end{tikzpicture}
	\caption{Percentage of false negatives for $width=8192$ and $th=0.1\%$.}
	\label{fig:hybrid6ev}
\end{subfigure}\hfill
\caption{Comparison of the Hybrid window approach for different depths and widths using a Count-Min sketch by classifying UDP and TCP packets using 10 different CAIDA traces. Confidence interval is 95\%.}\label{fig:hybrid}
\end{figure*}
By increasing the depth, the percentage of false positives decreases (Fig. \ref{fig:flushing1}, Fig. \ref{fig:flushing2} and Fig. \ref{fig:flushing3}). However, the percentage of false negatives increases (Fig. \ref{fig:flushing4}, Fig. \ref{fig:flushing5} and Fig. \ref{fig:flushing6}) and is higher than for all the other evaluated windowing approaches. This is expected, as the probability that one of the counts (from a set of them that are associated with the incoming packet, i.e. one per each stage) is reset to zero is higher. Since we use the Count-Min sketch in our experiments, the calculated minimum is below the threshold ($th\cdot N$). When an entry belonging to a heavy hitter flow is flushed, it will take at least $N\cdot th$ packets to reach the heavy hitter threshold again. In the meantime, the heavy hitter flow will not be identified as such. 

By decreasing the width, the number of false negatives reduces. The load factor per hash table is higher, and more collisions occur reducing the time needed for the counts to reach their previous value (Fig. \ref{fig:flushing6}).

\textbf{Hybrid window.} We have analyzed two different versions of the Hybrid Window: (1) without the initial ring data-structure of size $N/k$ and (2) with the initial ring data-structure. 

The accuracy of the first solution (without the initial ring data-structure) has lower accuracy than the analyzed ring window, but outperforms all the other analyzed solutions. The increased value of the false positives is due to the fact that smaller flows slowly accumulate in the counting sketch, before finally being removed after they hit $th \cdot N/m$. This probability is decreased with the increase of $th$, $N$ and width, similarly to the ring window (Fig. \ref{fig:hybrid1ev}, Fig. \ref{fig:hybrid2ev} and Fig. \ref{fig:hybrid3ev}). 

However, the advantage of this solution is that the probability of false negatives is equal to $0$. This is expected since the oldest batch of $th\cdot N /k$ packets is always removed from the second counting structure (that counts the number of $th\cdot N /k$ occurrences). As a consequence, the total count can only be overestimated, and never underestimated. 

To reduce the number of false positives, a smaller pure ring of size $N/m$ can be added to remove packets added to  the initial counting sketch (not from the $th\cdot N/m$ counter) after $N/m$ packets (second evaluated solution). The accuracy, with this initial ring data-structure, is comparable to the previously analyzed ring window. For a width of $8192$, the probability of false positives is under $2\%$ for all the analyzed values of $N$ and $th$ (Fig. \ref{fig:hybrid1ev} and Fig. \ref{fig:hybrid2ev}). Similarly to the ring window, the probability of false positives is reduced with the increase of the width (Fig. \ref{fig:hybrid3ev}). 

However, in contrast to the ring window, the probability of false negatives is no longer equal to 0\%.
False negatives occur when the initial ring window reduces the count for a heavy hitter flow after $N/k$ new packets, while the count remained under $th \cdot N/k$ and was never added to the second hash table (that counts the number of batches $th\cdot N /k$). However, this is corrected in the next few packets belonging to that flow and the probability of false negatives is smaller than $0.6\%$ for all the analyzed values of $th$, $width$ and $N$.

\subsection{Influence of parallel processing}
We evaluated our solutions using a Python program as well as by implementing the algorithm in P4 and testing it on a Netronome SmartNIC. We verified that our P4 code produced the same results as our Python implementation using multiple artificially generated packet traces by ensuring that the hash tables were identical in both cases at the end of the measurement interval and that the same packets were identified as belonging to a heavy hitter flow.

Afterwards, we evaluated our solutions using the CAIDA traces by generating packets at higher rates to evaluate the influence of parallel processing. Netronome SmartNICs process all incoming packets in parallel (60 microengines on Netronome cards in our testbed), leading to race conditions in cases when multiple microengines try to access the same register memory. The accuracy is decreased, and the probability of false positives and false negatives is increased. However, this difference is not significant (less than 0.5\%).

\section{Related work}\label{Sec_RelatedWork}
Calculating frequent items in a datastream is a well researched problem and many algorithms have been proposed over the years. We can divide them into three main groups: (1) sampling algorithms, (2) algorithms based on sketches, and (3) counting algorithms. 

\textbf{Sampling algorithms} (NetFlow \cite{netflow}, Sflow \cite{sflow}, Sample\&Hold \cite{Estan2002}) are currently widely deployed and used by network operators, but have some well-known limitations, like the trade-off between scalability, overhead, and accuracy. In these algorithms, nodes usually maintain current flow statistics that are periodically sent to a remote collecting point that performs detailed analysis. However, especially in core routers that process huge amounts of data, they can cause significant bandwidth, CPU, and memory overhead if the sampling rate is not set high enough \cite{netflowbanchmark}. Several modifications, that address these problems, were proposed in \cite{li2016, Zhu2015}. 

\textbf{Algorithms based on sketches} (Count-Min Sketch \cite{countmin}, UnivMon \cite{liu2016one}, Count Sketch \cite{count}, Probabilistic lossy counting \cite{Dimitropoulos2008}, CountMax \cite{yu2018countmax}, Elastic sketch \cite{yang2018elastic}) use specialized data structures called sketches that hash and count all packets in the switch hardware. They usually have lower memory requirements and can reduce processing time per packet, while processing every packet in a large stream of packets at the same time. However, they reduce accuracy causing potential overestimation or underestimation of the flow frequencies. Additionally, many of these algorithms (\cite{Einziger2016}) can not be easily and efficiently implemented on programmable switches (using languages such as P4 \cite{Bosshart2014}) and require specialized hardware. 

\textbf{Counting algorithms} (Hashpipe \cite{hashpipe}, Space-Saving Algorithm \cite{Metwally2005}, CSS \cite{ben2016heavy}) maintain a data-structure consisting only of heavy hitter flows and corresponding counts. Space-Saving algorithm is considered state-of-the-art in this group of algorithms as it has the lowest memory usage possible ($O(k)$) for a fixed accuracy among deterministic heavy hitter algorithms \cite{hashpipe, ben2016heavy}. It uses very simple actions (additions or subtractions), but requires either maintaining a sorted list or finding an item with the minimum counter value among all possible entries in the table. Unfortunately, both options are either not supported by existing programmable hardware or exceed the available processing budget. CSS improves the Space-Saving algorithm by using only statically allocated memory and by supporting constant time point queries \cite{ben2016heavy}. However, data-structures such as the TinyTable \cite{Einziger2016} were not developed with P4 in mind and cannot be efficiently maintained within the available time budget. Hashpipe \cite{hashpipe} uses a set of hash tables to count every packet received by the switch. It was designed for P4 and, as such, has very low processing overhead and memory consumption. 

\textbf{Sliding window approaches} (WCSS \cite{ben2016heavy}, SWAMP \cite{assaf2017,basat2017}, Memento \cite{basat2018memento}) remove the oldest entries from the counting data-structure so that only information about the last $N$ processed packets is present at the switch. SWAMP \cite{assaf2017,basat2017} maintains an additional array with flow identifiers from the last $N$ packets. Every time a new packet arrives the oldest entry from the array is removed and replaced with a new flow identifier. However, depending on the selected window size, memory consumption is very high. Ben-Basat et al. present two different solutions in \cite{ben2016heavy, basat2018memento} optimized for memory consumption with constant query time. However, maintaining data structures such as the TinyTable \cite{Einziger2016} as well as the presented sliding window structure requires many memory accesses exceeding the per-packet time budget and making it unsuitable for programmable network hardware.




\section{Conclusion}\label{Sec_Conclusion}
To avoid drops in throughput, programmable networking hardware comes with a set of specific constraints that need to be taken into account when designing new switch applications, such as a limited number of memory accesses as well as a very limited amount of memory to store stateful information. Most of the existing solutions to detect heavy hitters focus only on low memory overhead and do not take into account the limited number of memory accesses (typically just one read/modify/write per data-structure). To satisfy these new constraints, newer approaches, that maintain accuracy over time while having a low processing overhead and low memory consumption are needed.

By analyzing the Count-Min sketch, we realized that the memory used by the sketch could be distributed more optimally. All the packets were processed through a set of hash tables of the same size, leading to many collisions in all of them. This conclusion lead to the development of our own approach, the Gated Sketch. By using a set of hash tables, whose size decreases with the depth, and a set of thresholds, smaller flows are filtered in the first few tables. At the same time, as more memory is added to the first stages, the number of collisions is decreased. We have shown that our Gated sketch outperforms the Count-Min sketch with the same memory usage by lowering the number of false positives with a factor of 2.

Secondly, we focused on maintaining a high accuracy of the counting sketch over time without any intervention of the control-plane by using a sliding window implementation. We showed that the current approach (with the controller intervention), to flush the counting data-structure every $T$ seconds, has many drawbacks. For higher values of $T$, the percentage of false positives had a significant increase (up to $80\%$ of all packets processed by the switch). Additionally, on switches with high processing overhead, this approach required too many controller interventions making it unfeasible for smaller values of $N$ (window size). To counteract these issues, we developed and designed multiple different sliding window solutions and evaluated their accuracy for different values of $N$ and the heavy hitter threshold. We have shown that it is possible to maintain a high accuracy over time (Sequential Flushing and Hybrid Window) while taking into account the previously mentioned hardware requirements.

\bibliographystyle{acm}
\bibliography{ref}

\begin{thebibliography}{10}

\bibitem{caida2016trace}
The {CAIDA UCSD} anonymized internet traces - 2016, 2018.
\newblock Available at
  \url{http://www.caida.org/data/passive/passive_dataset.xml}.

\bibitem{assaf2017}
{\sc Assaf, E., Ben{-}Basat, R., Einziger, G., and Friedman, R.}
\newblock Pay for a sliding bloom filter and get counting, distinct elements,
  and entropy for free.
\newblock {\em CoRR abs/1712.01779\/} (2017).

\bibitem{basat2017}
{\sc Basat, R.~B., Einziger, G., Friedman, R., and Kassner, Y.}
\newblock Poster abstract: A sliding counting bloom filter.
\newblock In {\em 2017 IEEE Conference on Computer Communications Workshops
  (INFOCOM WKSHPS)\/} (May 2017), pp.~1012--1013.

\bibitem{basat2018memento}
{\sc Basat, R.~B., Einziger, G., Keslassy, I., Orda, A., Vargraftik, S., and
  Waisbard, E.}
\newblock Memento: Making sliding windows efficient for heavy hitters.
\newblock {\em arXiv preprint arXiv:1810.02899\/} (2018).

\bibitem{ben2016heavy}
{\sc Ben-Basat, R., Einziger, G., Friedman, R., and Kassner, Y.}
\newblock Heavy hitters in streams and sliding windows.
\newblock In {\em INFOCOM\/} (2016), pp.~1--9.

\bibitem{Benson2010}
{\sc Benson, T., Akella, A., and Maltz, D.~A.}
\newblock Network traffic characteristics of data centers in the wild.
\newblock In {\em Proceedings of the 10th ACM SIGCOMM Conference on Internet
  Measurement\/} (New York, NY, USA, 2010), IMC '10, ACM, pp.~267--280.

\bibitem{Bosshart2014}
{\sc Bosshart, P., Daly, D., Gibb, G., Izzard, M., McKeown, N., Rexford, J.,
  Schlesinger, C., Talayco, D., Vahdat, A., Varghese, G., and Walker, D.}
\newblock P4: programming protocol-independent packet processors.
\newblock {\em SIGCOMM Comput. Commun. Rev. 44}, 3 (2014), 87--95.

\bibitem{count}
{\sc Charikar, M., Chen, K., and Farach-Colton, M.}
\newblock Finding frequent items in data streams.
\newblock In {\em International Colloquium on Automata, Languages, and
  Programming\/} (2002), Springer, pp.~693--703.

\bibitem{netflowbanchmark}
{\sc Cisco~Systems, I.}
\newblock Netflow performance analysis.
\newblock Available at
  \url{https://www.cisco.com/c/dam/en/us/solutions/collateral/service-provider/secure-infrastructure/net_implementation_white_paper0900aecd80308a66.pdf}.

\bibitem{netflow}
{\sc Claise, B.}
\newblock {Cisco systems NetFlow services export version 9}.
\newblock Tech. Rep. 2070-1721, Internet Engineering Task Force, 2004.

\bibitem{countmin}
{\sc Cormode, G., and Muthukrishnan, S.}
\newblock An improved data stream summary: The count-min sketch and its
  applications.
\newblock {\em J. Algorithms 55}, 1 (Apr. 2005), 58--75.

\bibitem{Dimitropoulos2008}
{\sc Dimitropoulos, X., Hurley, P., and Kind, A.}
\newblock Probabilistic lossy counting: an efficient algorithm for finding
  heavy hitters.
\newblock {\em SIGCOMM Comput. Commun. Rev. 38}, 1 (2008), 5--5.

\bibitem{Einziger2016}
{\sc Einziger, G., and Friedman, R.}
\newblock Counting with tinytable: Every bit counts!
\newblock In {\em Proceedings of the 17th International Conference on
  Distributed Computing and Networking\/} (New York, NY, USA, 2016), ICDCN '16,
  ACM, pp.~27:1--27:10.

\bibitem{Estan2002}
{\sc Estan, C., and Varghese, G.}
\newblock New directions in traffic measurement and accounting.
\newblock {\em SIGCOMM Comput. Commun. Rev. 32}, 4 (Aug. 2002), 323--336.

\bibitem{kandula2009nature}
{\sc Kandula, S., Sengupta, S., Greenberg, A., Patel, P., and Chaiken, R.}
\newblock The nature of data center traffic: measurements \& analysis.
\newblock In {\em Proceedings of the 9th ACM SIGCOMM conference on Internet
  measurement\/} (2009), ACM, pp.~202--208.

\bibitem{li2016}
{\sc Li, Y., Miao, R., Kim, C., and Yu, M.}
\newblock Flowradar: A better netflow for data centers.
\newblock In {\em Nsdi\/} (2016), pp.~311--324.

\bibitem{liu2016one}
{\sc Liu, Z., Manousis, A., Vorsanger, G., Sekar, V., and Braverman, V.}
\newblock One sketch to rule them all: Rethinking network flow monitoring with
  univmon.
\newblock In {\em Proceedings of the 2016 ACM SIGCOMM Conference\/} (2016),
  ACM, pp.~101--114.

\bibitem{Metwally2005}
{\sc Metwally, A., Agrawal, D., and El~Abbadi, A.}
\newblock Efficient computation of frequent and top-k elements in data streams.
\newblock In {\em International Conference on Database Theory\/} (2005),
  Springer, pp.~398--412.

\bibitem{sflow}
{\sc Phaal, P., Panchen, S., and McKee, N.}
\newblock Inmon corporation's sflow: A method for monitoring traffic in
  switched and routed networks.
\newblock Tech. Rep. 2070-1721, Internet Engineering Task Force, 2001.

\bibitem{hashpipe}
{\sc Sivaraman, V., Narayana, S., Rottenstreich, O., Muthukrishnan, S., and
  Rexford, J.}
\newblock Heavy-hitter detection entirely in the data plane, 2017.

\bibitem{p4measurments}
{\sc Sonchack, J.}
\newblock {Feature Rich Flow Monitoring with. P4}.
\newblock Available at
  \url{https://www.netronome.com/media/documents/WBN-2017-11-1-Penn-Feature-Rich-Flow-Monitoring-OpenNFP\_.pdf}.

\bibitem{belma}
{\sc Turkovic, B., Kuipers, F., van Adrichem, N., and Langendoen, K.}
\newblock Fast network congestion detection and avoidance using p4.
\newblock In {\em Proceedings of the 2018 Workshop on Networking for Emerging
  Applications and Technologies\/} (New York, NY, USA, 2018), NEAT '18, ACM,
  pp.~45--51.

\bibitem{yang2018elastic}
{\sc Yang, T., Jiang, J., Liu, P., Huang, Q., Gong, J., Zhou, Y., Miao, R., Li,
  X., and Uhlig, S.}
\newblock Elastic sketch: Adaptive and fast network-wide measurements.
\newblock In {\em Proceedings of the 2018 Conference of the ACM Special
  Interest Group on Data Communication\/} (2018), ACM, pp.~561--575.

\bibitem{yu2018countmax}
{\sc Yu, X., Xu, H., Yao, D., Wang, H., and Huang, L.}
\newblock Countmax: A lightweight and cooperative sketch measurement for
  software-defined networks.
\newblock {\em IEEE/ACM Transactions on Networking\/} (2018).

\bibitem{Zhu2015}
{\sc Zhu, Y., Kang, N., Cao, J., Greenberg, A., Lu, G., Mahajan, R., Maltz, D.,
  Yuan, L., Zhang, M., Zhao, B.~Y., and Zheng, H.}
\newblock Packet-level telemetry in large datacenter networks.
\newblock {\em SIGCOMM Comput. Commun. Rev. 45}, 4 (2015), 479--491.

\end{thebibliography}

\end{document}